\documentclass{article}

\usepackage{soul}

\usepackage[utf8]{inputenc}
\usepackage{graphicx} % Required for inserting images
\usepackage[top = 1in, bottom = 1 in , left= 1 in, right = 0.5 in]{geometry}
\usepackage{amsmath,bm}
\usepackage[title]{appendix}
\usepackage{authblk}
\usepackage{xcolor}
\usepackage{comment}
\usepackage{gensymb}

\title{Quasi-BICs due to symmetry mismatch  in  architected elastic plates}
\author[1]{Adib Rahman}
\author[2]{Sean Perkins}
\author[1]{Raj Kumar Pal}
\affil[1]{Department of Mechanical Engineering, Texas A\&M University, College Station, Texas 77843, USA}
\affil[2]{Department of Mechanical and Nuclear Engineering, Kansas State University, Manhattan, Kansas 66506, USA}
\date{}

\begin{document}

\maketitle
\begin{abstract}

%%%%% note to write a abstract

% 1.	What is the problem you are solving? 
% 2.	How is it different from previous works?  
% 3.	What is your approach? 
% 4.	What are your key results? 
% 5.	What are the implications? 

We report the existence of quasi-bound modes in the continuum (quasi-BICs) in architected elastic plates based on a square lattice. The structure consists of topologically trivial and nontrivial lattices, forming an interface and maintaining $C_2$ symmetry. Carefully engineered interface gives rise to center quasi-BICs. We show how the mismatch in symmetry between the Bloch modes of the lattice and the defect modes confine them at the defect center, resulting in  quasi-BICs. Our analysis begins with a square lattice-based spring-mass system. Finite element simulations on an architected plate comprising of slender curved and straight beams to achieve the desired stiffness variation predict quasi-BICs analogous to those in the discrete model. These predictions are validated with Laser Doppler vibrometry based experiments,  confirming the presence of a quasi-BIC in the structure. The concept of quasi-BICs arising from modal symmetry mismatch paves the way for achieving localized modes in elastic structures, with potential applications as resonators. 
\end{abstract}

\section{Introduction}\label{sec:intro}
% Points:
% \begin{itemize}
%    \item talk about localized mode
%    \item what people have done widely
%    \item What is the status of works in elastic field
%    \item what is the problem in elastic field
%    \item what we have done
% \end{itemize}

Architected structures or meta-structures have unique wave manipulation capabilities, including localization, negative refraction, and unidirectional propagation~\cite{hussein2014dynamics,hernandez2008localized,zhu2014negative,zanotto2022metamaterial}. These wave phenomena are achieved by designing geometric features of the structures. However, most such phenomena rely on the existence of a frequency bandgap~\cite{li2012elastic,jui2024topological}. Achieving a bandgap in a desired frequency range can be challenging in a structure due to space (in Bragg bandgaps) or bandwidth (in local-resonance bandgaps) limitations. In this context, bound modes/states in the continuum (BICs) present a promising alternative, offering wave confinement without requiring a frequency band gap~\cite{hsu2016bound}. The amplitude of a BIC goes to zero outside a compact region in space, ensuring zero energy leakage to surroundings and thereby giving a route to achieve extremely high quality factors. Recent advances in manufacturing have accelerated the search for BICs across physical domains, particularly in photonics~\cite{xu2023recent} and acoustics~\cite{huang2022general}. BICs have potential applications in lasing~\cite{imada1999coherent,hirose2014watt,kodigala2017lasing}, sensing~\cite{Romano:19}, filtering~\cite{Doskolovich:19,foley2014symmetry}, supersonic surface acoustic device~\cite{kawachi2001optimal,naumenko2003surface}, vibration absorption~\cite{cao2021perfect}, and wave guiding~\cite{benabid2002stimulated,couny2007generation,rahman2022bound}. 

BICs generally fall into two major categories: symmetry-protected BICs and accidental BICs. Symmetry-protected BICs arise when symmetry of a mode shape is distinct from symmetry of the supporting structure. Because of the symmetry mismatch, the mode shape does not hybridize with other bulk modes at nearby frequencies and stays confined. This phenomena can be achieved by eliminating a particular symmetric feature in a specific direction, while maintaining other symmetries of the structure~\cite{PhysRevApplied.19.054001} or by making the system asymmetric in every direction~\cite{cong2019symmetry,PhysRevA.100.063803}. Such BICs have been experimentally observed in various systems, such as periodic chains of dielectric disks~\cite{sadrieva2019experimental}, and optical waveguides~\cite{plotnik2011experimental}. In contrast, accidental BICs originate from phenomena like destructive interference of reflected waves from the far field~\cite{PhysRevLett.100.183902,sadreev2021interference} or precisely engineered defects in the structure~\cite{bulgakov2008bound,marinica2008bound,vaidya2021point}. Recently, topologically protected BICs have been achieved in photonic crystal slabs~\cite{zhen2014topological} and higher-order topological insulators by exploiting specific symmetries of the structures~\cite{benalcazar2020bound,cerjan2020observation,wang2021quantum}, with topological properties guaranteeing the existence of BICs. Despite these noteworthy studies, realizing BICs is challenging in general due to stringent dimension requirements and sensitivity to manufacturing imperfections. An alternative concept with less stringent requirements, called quasi-bound modes in the continuum (quasi-BICs or Q-BICs) has been proposed~\cite{zhang2023super,farhat2024observation,overvig2021chiral,huang2022moire} in recent years. Quasi-BICs are leaky resonances or decaying localized modes whose frequencies lie in the  passband. They typically offer high energy confinement in small regions~\cite{vial2024platonic}. 

Achieving elastic BICs is challenging due to the simultaneous presence of multiple types of traveling waves, including shear, longitudinal and flexural. A BIC requires that there is no coupling between a localized mode shape and all these  traveling waves~\cite{rahman2024elastic}. The resulting large number of constraints make designs complex or infeasible. Most studies to date have focused on achieving BICs in effectively one-dimensional ($1D$) structures. Theoretical predictions of elastic BICs were first made by Haq and Shabanov for in-plane waves in a plate with scatterers~\cite{haq2021bound}. Notable demonstrations of quasi-BICs have been achieved at the boundaries of elastic plates with arrays of resonators~\cite{cao2021perfect,cao2021elastic}, in architected beams~\cite{rahman2024elastic} and in waveguides attached to an array of reflection-symmetric pillar resonators~\cite{an2024multibranch}. There have also been works exploiting non-hermiticity (damping) and boundary conditions to achieve BICs at the edges of a periodic structure~\cite{fan2022observation}. 
In contrast, BICs in two-dimensional ($2D$) structures are limited. An example is the recent demonstration of quasi-BICs in a silicon plate with silica defect~\cite{gao2024bound}. In addition, periodic structures with higher order topological properties are emerging as potential candidates to achieve corner localized mode in $2D$ elastic  structures~\cite{fan2019elastic,zheng2022higher,chen2021corner}, although the corner localized modes are found in frequency bandgap. A challenge with corner localized modes in passband is that their frequencies are degenerate with those of multiple bulk modes~\cite{benalcazar2020bound}, making them hard to realize experimentally.  

% what gaps is the current work addressing and how: 
Here, we seek to realize quasi-BICs at point defects in architected plates whose frequencies are not degenerate with bulk modes. To this end, we consider a lattice comprising of two types of square lattices whose unit cells are related by a translation. As a result, it does not have the four fold rotation  ($C_4$) symmetry of the constituent square lattices, which enables breaking the degeneracy of bulk modes at the localized mode frequency. We use a discrete spring mass system to predict the existence of center-localized quasi-BICs. After that, we build a $3D$ model that mimics the similar stiffness and mass distribution in the spring mass system. We then conduct numerical simulations and experiments to verify and validate the existence of center localized quasi-BICs in the structure. 

The outline of this paper is as follows: Sec.~\ref{sec:theory} presents the theory of localized modes at the interface between lattices using a discrete spring mass system. In Sec.~\ref{sec:Numerical_and_exp}, we present the design of an architected plate, accompanied by numerical simulations and experiments that validate the existence of quasi-BICs. Finally the conclusions are summarized in Sec.~\ref{sec:con}. 

\section{Theory using discrete spring-mass lattices}\label{sec:theory}
We start by considering discrete spring mass lattices to explain the underlying ideas and aid the design of an architected plate. We first present a brief review of a $C_4$ symmetric square lattice that supports topological corner localized modes in frequency pass band. These localized modes are not easily observable since there are multiple bulk modes with the same frequency. To break this frequency degeneracy, we introduce interfaces between two types of lattices that are shifted copies of each other. The resulting lattice, now $C_2$ symmetric, supports defect modes at the center with frequencies in the pass band. We discuss how the mismatch in symmetry between traveling Bloch wave and the defect modes results in their being strongly localized at the defect core.

\subsection{Brief review of topological corner localized modes in square lattices}\label{sec:brief_square}

\begin{figure}[hbtp!]
    \centering
    \includegraphics[width=15cm]{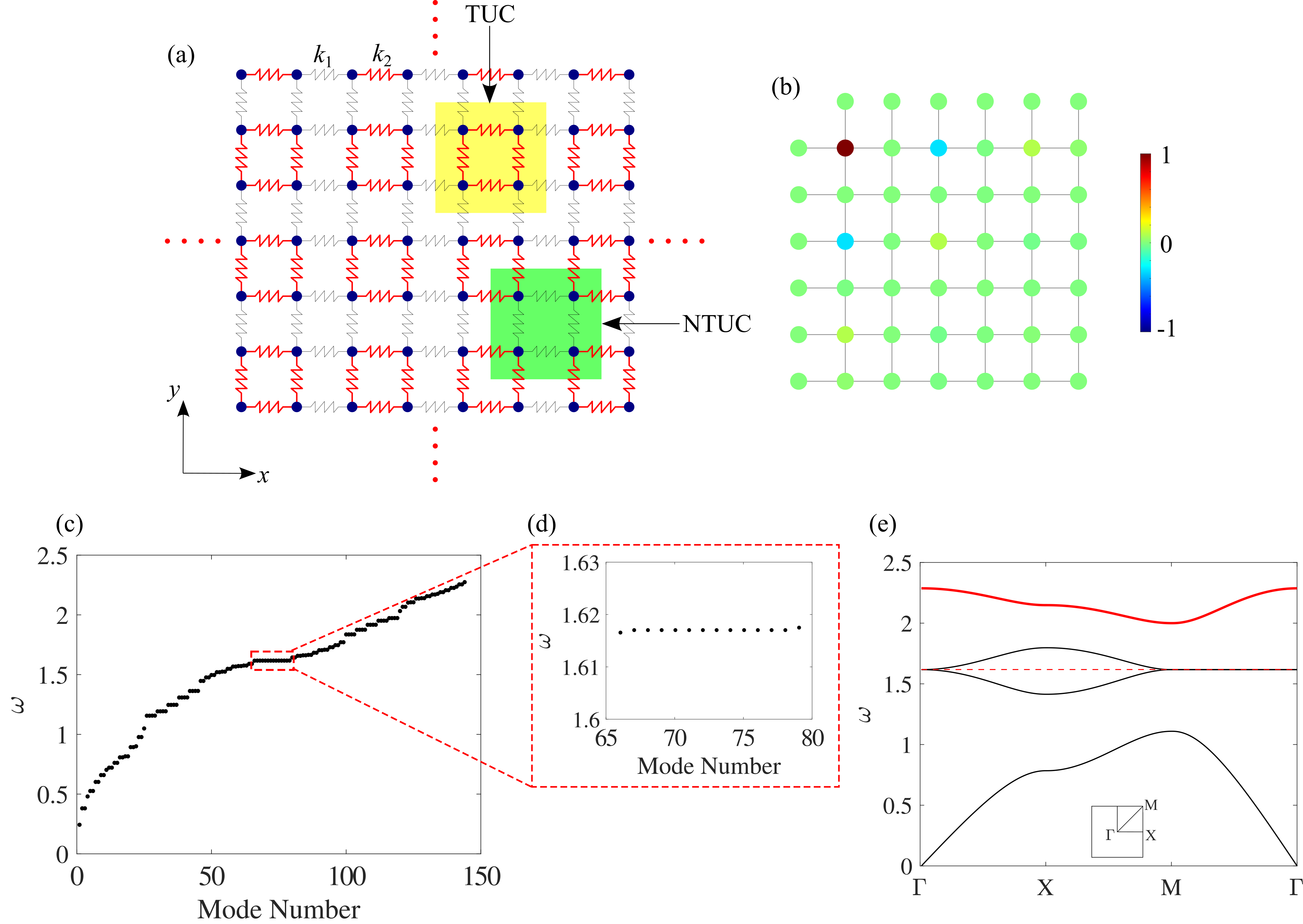}
    \caption{(a) Schematic of  square lattice. Identical masses connected by red/darker $k_2$ springs and black/lighter $k_1$ springs, $k_1 < k_2$. Trivial (TUC) and  nontrivial (NTUC) unit cells are shaded. (b) Corner localized mode at top left corner of a finite lattice of NTUC. (c) Natural frequencies of a lattice of $6\times6$ NTUC. (d) Zoomed in view of red boxed region in (c) shows multiple bulk modes are degenerate at the corner mode frequency $\omega =1.6186$. (e) Dispersion curves of the lattice over irreducible Brillouin zone (IBZ) boundary $\Gamma XM\Gamma$, where red dashed line marks location of frequency of corner mode shown in (b).}
    \label{fig:square_lat}
\end{figure}

Figure~\ref{fig:square_lat}(a) displays a periodic lattice comprising of discrete springs and masses. All the nodes have identical masses of unit magnitude, \textit{i.e.,}  $m=1$. Each mass has one degree of freedom and can move out-of-plane. The nodal masses are connected by springs of two different stiffness: $k_1$ (black/thin lined springs) and $k_2$ (red/thick lined springs), with $k_2 > k_1$. The lattice presents four distinct choices of unit cells that are translated copies of each other. The yellow and green shaded boxes in Fig.~\ref{fig:square_lat}(a) show two of these choices. Both unit cells have four masses, but differ in their spring connectivity. The springs within a unit cell and those connecting masses lying in adjacent unit cells are hereby referred to as intra- and inter-unit cell springs, respectively. In the unit cell within the yellow shaded box, the intra- and inter-unit cell springs have stiffness $k_2$ and $k_1$, respectively. In contrast, the intra- and inter-unit cell spring stiffness values in the  green shaded box are reversed. This lattice is inspired by the design studied in~\cite{benalcazar2020bound} for electron states in atomic lattices. 

The dispersion surface of lattices generated with unit cells in yellow and green boxes have distinct higher order topological invariants~\cite{benalcazar2020bound,PhysRevB.100.075120}. The choice of unit cell thus dictates if a lattice will be topologically trivial or nontrivial. In particular, a lattice will be nontrivial (trivial) when its inter-unit cell stiffness is higher (lower) than intra-unit cell stiffness. Studies in quantum~\cite{benalcazar2020bound} and acoustic~\cite{PhysRevB.100.075120,wu2023square} systems have shown that a finite square lattice comprising of an integer number of nontrivial unit cells supports localized modes at its corners. Note that the infinite lattices generated with trivial or nontrivial unit cells are the same. Corner localized modes are thus present or absent depending on how an infinite lattice is truncated to generate the finite lattice.

\begin{figure}[!b]
    \centering
    \includegraphics[width = 16cm]{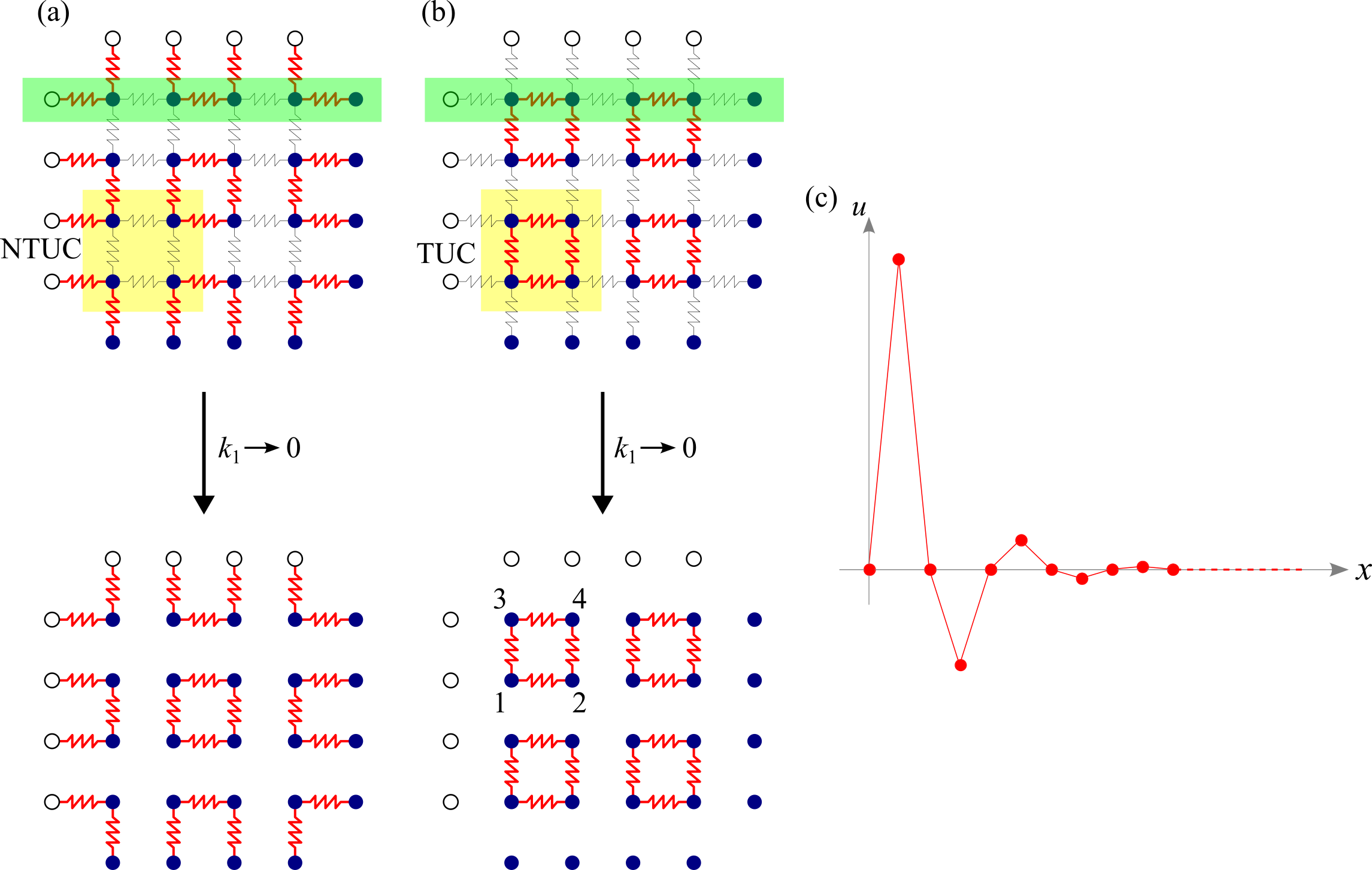}
    \caption{Top left corners of lattices comprising of (a)nontrivial and (b) trivial unit cells along with the resulting lattices in the limit $k_1 \to 0$. Top rows marked by green sheds are analogous to (a)nontrivial and (b)trivial SSH chains. (c) Corner localized mode's displacement along the top row of (a).  }
    \label{fig:zero_corner}
\end{figure}

We determined the mode shapes of finite trivial and nontrivial lattices comprising of $6 \times 6$ unit cells. Figure~\ref{fig:zero_corner}(a,b) display the top-left corners of both these lattices. The blue nodes are free and the nodes at the fixed boundaries are marked by void circles. We point to two key observations regarding the mode shapes of these lattices. First, corner localized modes arise only in the nontrivial lattice. Figure~\ref{fig:square_lat}(b) displays the mode localized at this corner in the nontrivial lattice. Second, in a large nontrivial lattice, multiple bulk modes appear at same corner mode frequency. Figure~\ref{fig:square_lat}(c) displays the natural frequencies of the finite nontrivial lattice. A zoomed-in view of the red boxed region in Fig.~\ref{fig:square_lat}(c) in Fig.~\ref{fig:square_lat}(d) shows that several globally spanning bulk modes coexist at the corner mode frequency. This degeneracy causes the corner mode to hybridize with the bulk modes, making the former difficult to isolate. We give heuristic explanations below for these two observations, a more rigorous analysis is presented in~\cite{benalcazar2020bound,PhysRevB.100.075120}. 

Now, to understand why only the nontrivial lattice supports corner localized modes, let us focus on their respective top rows, i.e., the spring-mass chains shaded by green in Fig.~\ref{fig:zero_corner}(a) and Fig.~\ref{fig:zero_corner}(b). In the nontrivial lattice, this chain is a nontrivial SSH chain. The springs connecting adjacent masses have alternating stiffness. In addition to each mass having an identical spring connecting it with the next layer, which may be viewed as a ground spring for this chain. Note that each of the four edges has the same nontrivial SSH lattice. Its ends support topological localized modes\cite{padavic2018topological}, which manifest as corner modes in the square lattice. On the other hand, the chain at any edge of the trivial lattice corresponds to the trivial SSH chain, which does not support any localized modes. The natural frequency and decay rate of the localized mode can be explicitly determined. 
Figure~\ref{fig:zero_corner}(c) displays how the mode decays as we move from the left end to the right in the green shaded chain. Each alternating mass is at rest. Let $u_p$ denote the displacement the $p$-th mass from the left end in this chain. The index $p$ starts from zero at the fixed node (unfilled circle in the schematic). The governing equation for the first mass is thus $m\ddot{u}_1 + 2(k_1+k_2)u_1 = 0$, and the corner localized mode frequency is $\omega_c = \sqrt{2(k_1 + k_2)/m}$. Similarly, the governing equation for the second mass is $m \ddot{u}_2 + k_1 (u_2 - u_1) + k_2(u_2-u_3) = 0$. Since it is at rest, the equation reduces to $k_1 u_1  + k_2 u_3 = 0$, which then gives the decay rate $|u_3/u_1| = k_1/k_2 < 1$. The numerical results are presented for $k_2=1$ and $k_1/k_2=0.3076$ as this choice of stiffness ratio matches with experimental parameters, discussed later in Sec.~\ref{sec:3d_design}. The natural frequency is $\omega_c$ is 1.6186. Figure~\ref{fig:square_lat}(e) shows the dispersion diagram of this lattice along with the localized mode frequency in 
a dashed line. When the boundary nodes are fixed, the corner mode frequency lies at intersection of second and third frequency bands for any $k_1 < k_2$ For other boundary conditions, this frequency may lie in the  bandgap.

\begin{figure}[!b]
    \centering
    \includegraphics[width = 14cm]{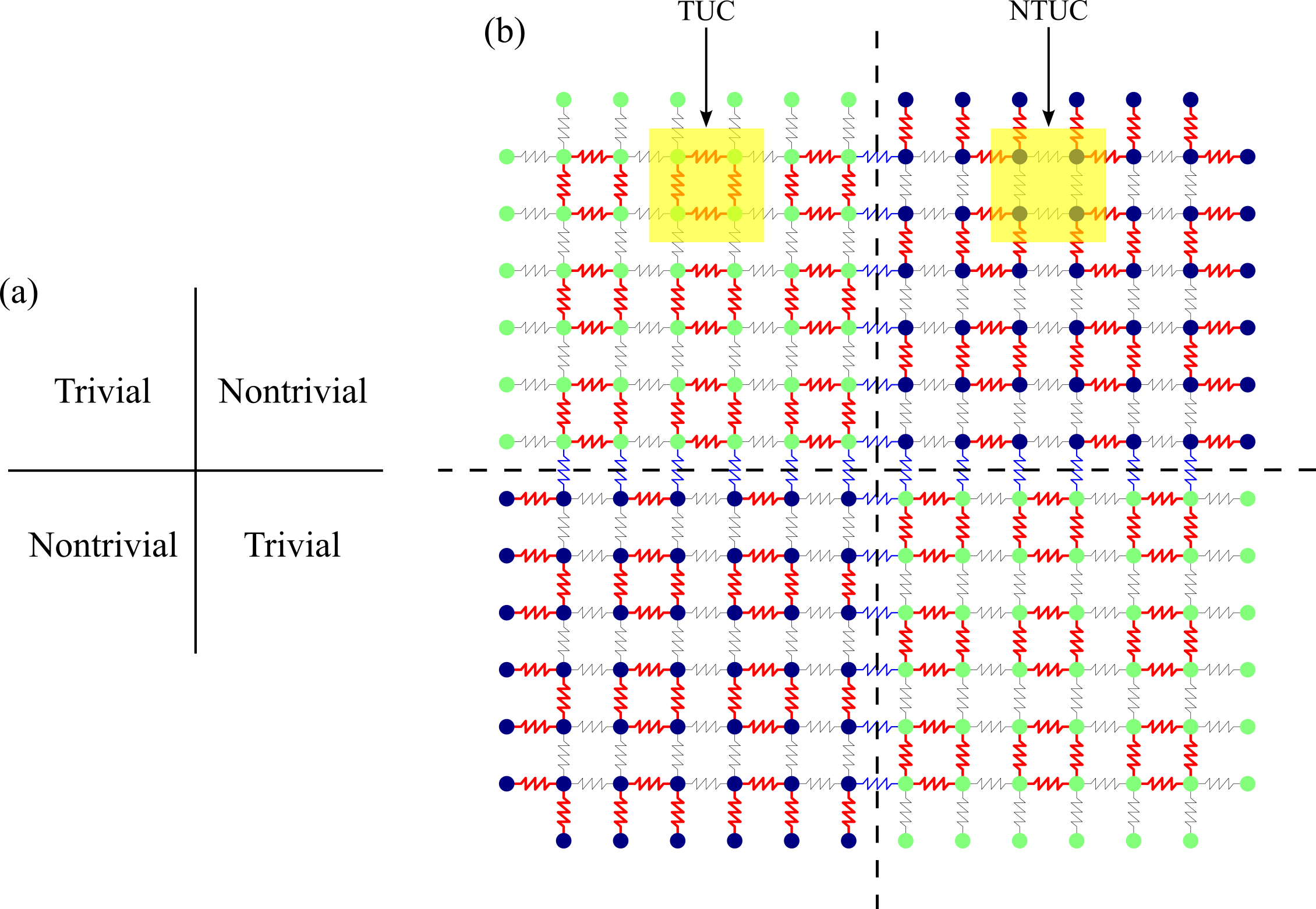}
    \caption{Proposed lattice design. Two trivial and two nontrivial square lattices are combined as shown in (a). The interface is denoted by dashed lines in (b). Interface springs (blue) have stiffness $k_i$. Shaded box on the left (right) indicates a trivial (nontrivial) unit cell. This structure now has $C_2$ symmetry instead of $C_4$. }
    \label{fig:spring_lat}
\end{figure}

Next, let us give an intuitive explanation for the presence of multiple degenerate modes with same frequency as the corner localized mode using the limiting case $k_1 \ll k_2$. Fixing $k_2=1$ and setting $k_1=0$, the corner of the finite nontrivial lattice reduces to the one displayed in Fig.~\ref{fig:zero_corner}(a). The unit cells, both in the interior and along the boundaries, are now isolated from each other. Since they are isolated, their natural frequencies can be determined explicitly. The corner localized mode arises in this limiting case as the corner node is connected to two fixed boundary nodes by stiffness $k_2$ (see Fig.~\ref{fig:zero_corner}(a). Its frequency is $\sqrt{2k_2/m}$. Similarly, the frequencies of each of the interior isolated unit cells are $\omega = \{ 0,\sqrt{2k_2/m}, \sqrt{2k_2/m}, 2\sqrt{k_2/m} \}$. Thus there are two modes in each of the interior unit cells with the same frequency as the corner localized mode, with mode shapes $(u_1,u_2,u_3,u_4) = (1,0,0,-1)$ and $(0,-1,1,0)$. Here, the nodal indices are as marked in the bottom panel in Fig.~\ref{fig:spring_lat}b. As mentioned above, the combination of $C_4$ symmetry and the absence of diagonal springs leads to this degeneracy between bulk and corner mode frequencies for any stiffness $k_1$ satisfying $k_1 < k_2$. In the following sections, we show how breaking $C_4$ symmetry of the finite lattice removes this degeneracy. 

\subsection{Proposed lattice design and its mode shapes} \label{sec:prop_lat}

Although the lattice of nontrivial unit cells in Sec.~\ref{sec:brief_square} supports corner localized modes in pass band, it has two obstacles making their realization challenging in continuous elastic media. The first is the degeneracy with  multiple bulk mode frequencies. The second is that fixed boundary conditions are hard to enforce as support structures are not infinitely rigid. Here we propose a design that overcomes these obstacles by instead achieving quasi-BICs in the interior of a lattice. 

\subsubsection{Lattice with interfaces to break $C_4$ symmetry}\label{sec:lat_int}

Figure~\ref{fig:spring_lat}(b) displays a schematic of the lattice design. It comprises of four copies of square lattices with interfaces between them. We combine two $n\times n$ trivial and two $n\times n$ nontrivial lattices ($n=3$ in the figure) as shown in Fig.~\ref{fig:spring_lat}(b). Unit cells for both trivial and nontrivial lattices are marked by shaded boxes. The resulting lattice no longer has $C_4$ symmetry, but now has $C_2$ symmetry. Dashed lines show the boundary between the four segments. The red (thick) and black (thin) springs have high ($k_2$) and low ($k_1$) stiffness values, respectively. Springs at the interface (blue color) of the trivial and nontrivial lattices have a different stiffness $k_i$. 

We set $k_1 = 0.3076$, $k_2=1$ and $k_i=2$ for numerical simulations. As mentioned above, these choices are made so that the stiffness ratios match with those in the fabricated elastic lattice discussed later in Sec.~\ref{sec:Numerical_and_exp}. This choice of $k_i$ ensures the center localized mode lies in the frequency passband and its frequency is not degenerate with that of the corner modes. In Appendix~\ref{sec:centerModeVary}, we show  the effect of interface stiffness $k_i$ on the center localized mode's frequency.  Although the results are presented here for a lattice having $3\times3$  unit cells in each of the 4 segments, the concepts translate to larger lattice sizes. 

\subsubsection{Localized mode shapes of the lattice} \label{sec:localized_modes}

\begin{figure}[!b]
    \centering
    \includegraphics[height = 6cm]{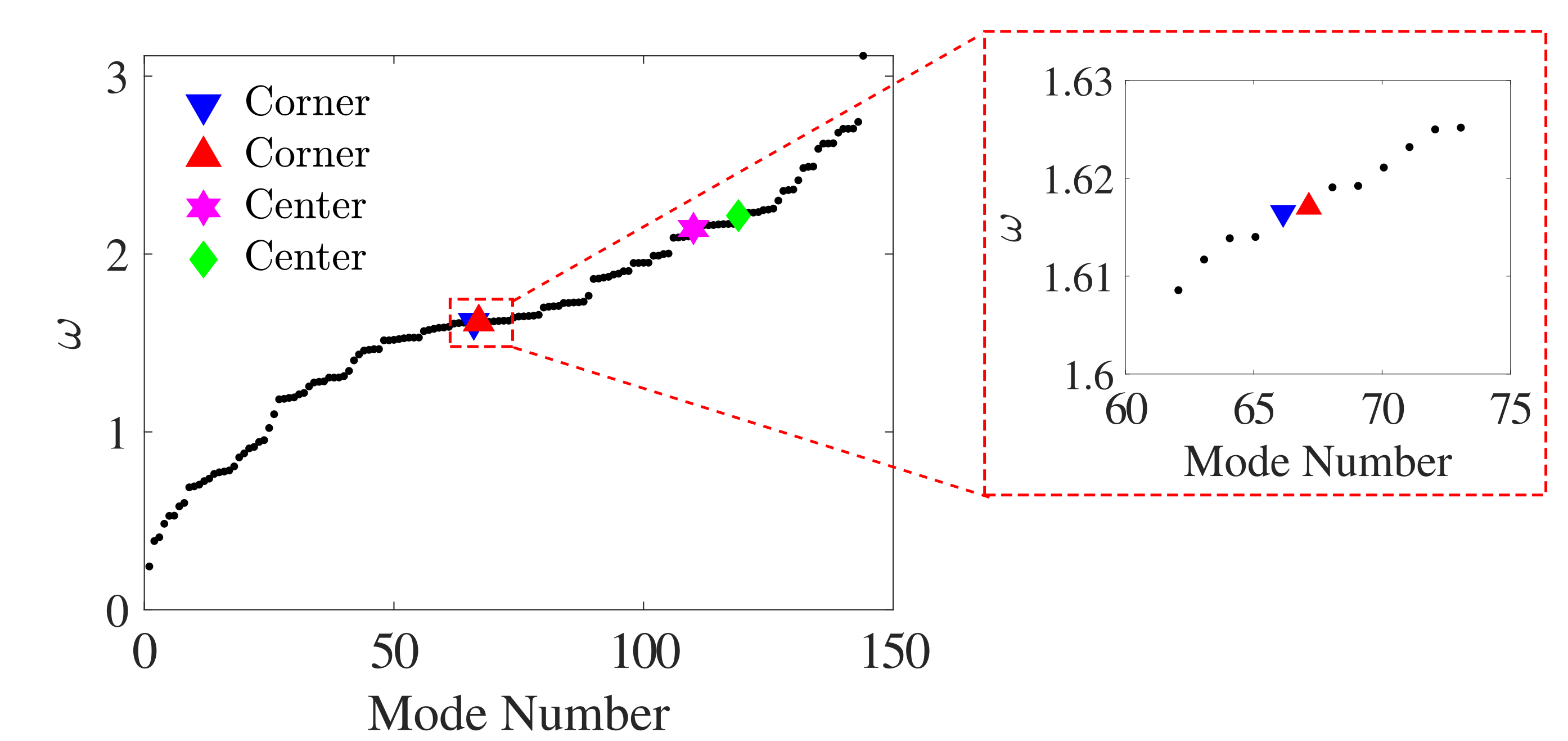}
    \caption{(a) Natural frequencies of the lattice shown in Fig.~\ref{fig:spring_lat}(b). Center and corner localized mode frequencies that lie in a pass band are indicated by large markers. (b) Zoomed in view of red boxed region in (a) shows there is no degeneracy of the two corner modes with bulk mode frequencies. }
    \label{fig:spr_nat_frq}
\end{figure}

Let us examine the natural frequencies and mode shapes of the finite lattice in Fig.~\ref{fig:spring_lat}(b). There are $3\times 3\times 4 = 36$ masses in each segment and thus $36\times4 = 144$ masses in the finite lattice. In addition, a layer of fixed nodes is added on all four boundaries. The natural frequencies are presented in Fig.~\ref{fig:spr_nat_frq}(a), where black dots indicate modes spanning the lattice while the four larger colored markers indicate corner and center localized modes. See Appendix~\ref{sec:nat_freq} for a brief procedure on determining the frequencies and mode shapes. A zoomed in view of the region enclosed by the red box in Fig.~\ref{fig:spr_nat_frq}(a) is shown in Fig.~\ref{fig:spr_nat_frq}(b), which clearly illustrates the absence of other bulk modes at the corner mode frequencies that are marked by red and blue triangles.  In contrast, recall that Fig.~\ref{fig:square_lat}(d) shows the degeneracy at corner mode frequency when the lattice comprises entirely of nontrivial unit cells. The proposed design thus demonstrates the effectiveness of our approach in eliminating the degeneracy. 

\begin{figure}[!t]
    \centering
    \includegraphics[height = 6cm]{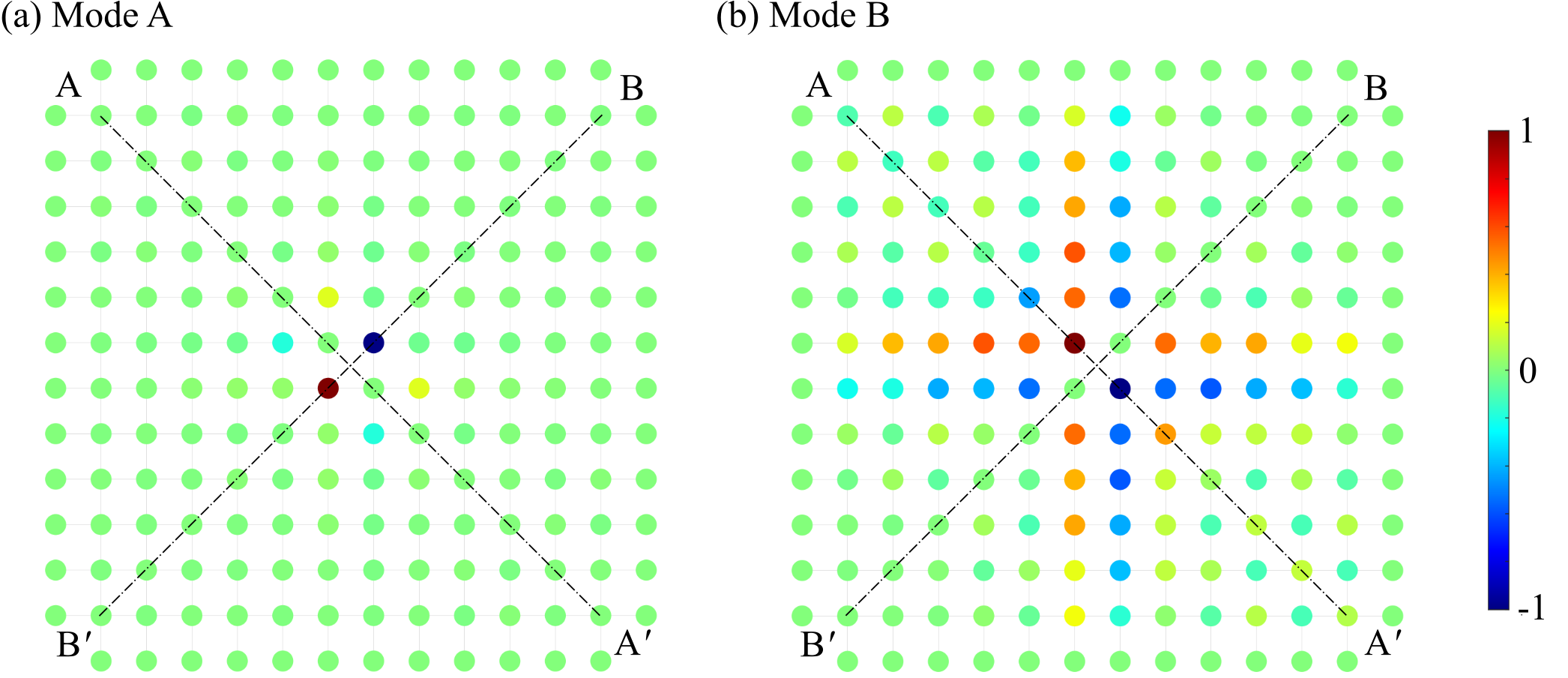}
    \caption{Mode shapes of the two center quasi-BICs. The mode shapes in (a) and (b) are anti-symmetric about the $AA^\prime$ and $BB^\prime$ diagonals, respectively. }
    \label{fig:spring_mode}
\end{figure}

The center localized mode shapes whose frequencies are marked by magenta star and green diamond in Fig.~\ref{fig:spr_nat_frq}(a) are shown in Fig.~\ref{fig:spring_mode}(a) and ~\ref{fig:spring_mode}(b), respectively. Hereafter, we focus on these center localized modes in the rest of this article and will not discuss corner localized modes that are also supported that the top-right and bottom-left corners of this lattice. To check if these center localized modes are BICs, we examine if their frequencies lie in the frequency passband of dispersion surface. Figure~\ref{fig:square_lat}(e) displays the dispersion curves along the high-symmetry path $\Gamma X M \Gamma$ at the boundary of the irreducible Brillouin zone (IBZ) for the lattice. See Appendix~\ref{sec:disp_analysis} for details of dispersion calculation procedure. We find that the localized modes illustrated in Fig.~\ref{fig:spring_mode} lie in the frequency branch marked by red line in Fig.~\ref{fig:square_lat}(e) but do not qualify as BICs since they are not completely leak-free, \textit{i.e.,} they are quasi-BICs.

\begin{figure}[!b]
    \centering
    \includegraphics[height = 11cm]{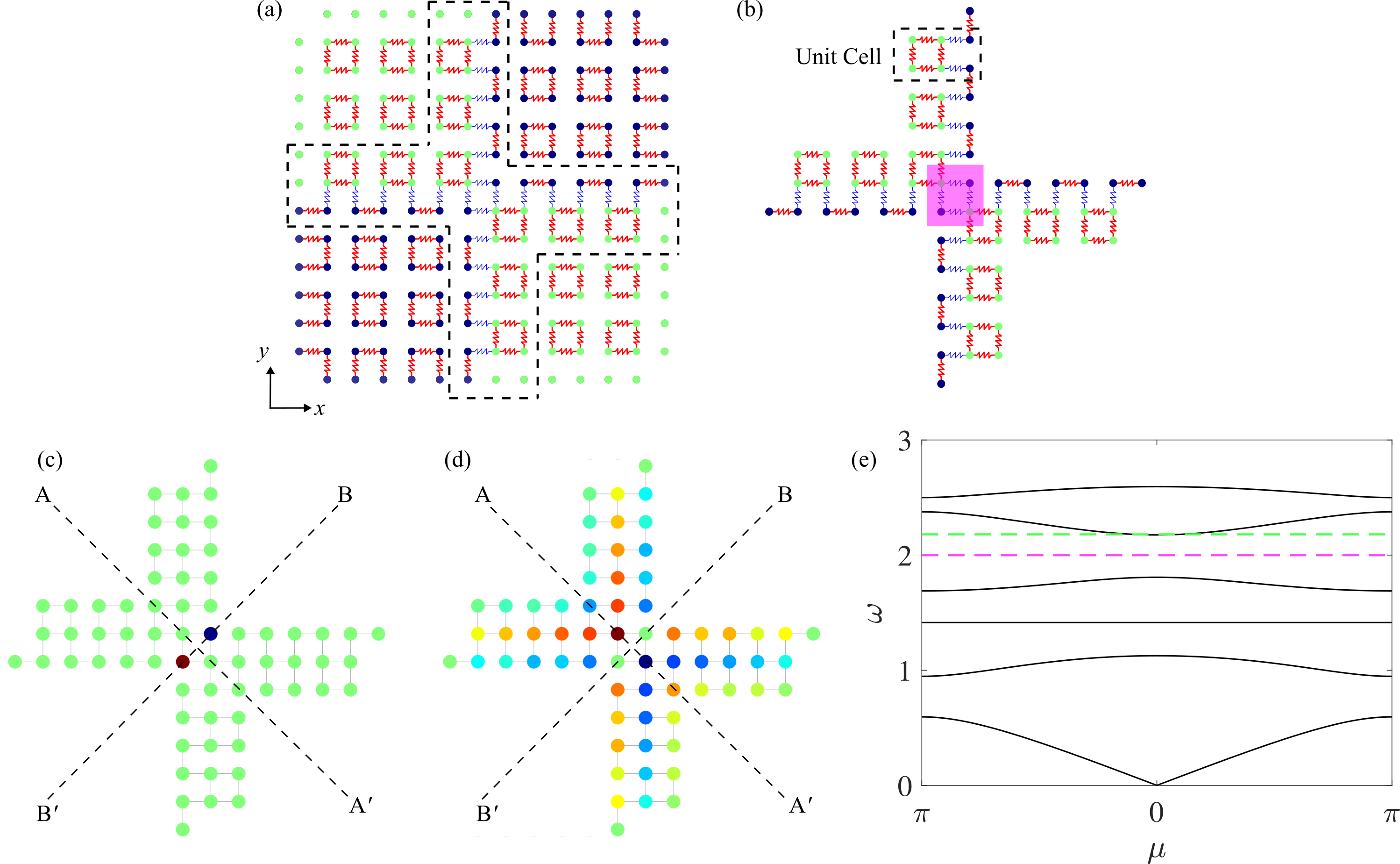}
    \caption{(a) Lattice shown in the Fig.~\ref{fig:spring_lat}(b) in the limit of spring stiffness $k_1 \rightarrow 0$. Center block enclosed by dashed lines contains the interface. It becomes disjoint from the remaining structure and is shown separately in (b). (c,d) Localized modes in this center block. (e) Dispersion diagram of the periodic chain in (b). Magenta and green dashed lines indicate  frequency of modes in (c) and (d), respectively.}
    \label{fig:zero_main}
\end{figure}

\subsection{Center localized modes: reasons for their decay}\label{sec:dis_reason}
Having shown the presence of center quasi-BICs, let us now analyze the reasons for their decay away from the defect center. We do this in two parts. First, we show that these modes decay along the interface since they lie in the frequency bandgap of the spring-mass chain along the interface. Next, we show that these modes decay in the interior by illustrating their mismatch in symmetry with the traveling Bloch modes of both nontrivial and trivial lattices.

\subsubsection{Dispersion analysis of interface chain}\label{sec:chain_disp}

Again, to gain insight, we analyze the lattice in the limiting case of $k_1\rightarrow0$. The resulting lattice is shown in Fig.~\ref{fig:zero_main}(a), where the region enclosed by black dashed lines is a monolithic block with no connection with rest of the lattice. This block is shown separately in Fig.~\ref{fig:zero_main}(b). This block supports the center localized modes displayed in Fig.~\ref{fig:zero_main}(c,d) that are analogues of the entire lattice's center quasi-BICs. Let us analyze the dynamics of this block to understand why the mode decays along it. 

In this block, the two interface chains are identical and they are periodic along the $x$ and $y$ directions, respectively. Their unit cell, shown in the dashed box in Fig.~\ref{fig:zero_main}(b), has six nodes. The shaded region in Fig.~\ref{fig:zero_main}(b) where the two chains meet may be viewed as a defect in the periodic chain. The mode localized at the defect decays away from it along this periodic structure as its frequency lies in the bandgap of this periodic chain's dispersion surface. The dispersion curves of the $1D$ periodic structure are shown in Fig.~\ref{fig:zero_main}(e), where $\mu$ is non-dimensional wavenumber.

In the dispersion diagram of the interface chain shown in Fig.~\ref{fig:zero_main}(e), the frequencies of the localized mode in Fig.~\ref{fig:zero_main}(c) and~\ref{fig:zero_main}(d) are indicated by magenta and green dashed lines, respectively. The magenta dashed line is completely in band gap. Thus, the mode in Fig.~\ref{fig:zero_main}(c) is strongly localized at defect center and shows rapid decay along the interface. On the other hand, the frequency of the mode in Fig.~\ref{fig:zero_main}(d) just lies outside the pass band and thus results in a slowly decaying mode. Now, instead of zero $k_1$, the lattice with $k_1 = 0.3076$ supports similar modes decaying along the interface shown in Fig.~\ref{fig:spring_mode}.

Let us remark on how the mode shape of the center localized mode in  Fig.~\ref{fig:zero_main}(c) is similar to the corner mode discussed earlier in Sec.~\ref{sec:brief_square}. They exhibit same wave decay pattern in the nontrivial segments. Note from Fig.~\ref{fig:zero_main}(c) that the center nodes along diagonal $BB^\prime$ are connected only to the center nodes along the diagonal $AA^\prime$ via springs of stiffness $k_i$. An equal and out-of-phase motion of the nodes along $BB^\prime$ kepng the center nodes along $AA^\prime$ at rest. Now, if we have a nonzero $k_1$ as in the full lattice, the center localized mode will decay along edges connected to the out-of-phase moving center nodes. The decay pattern is same as that shown in Fig.~\ref{fig:zero_corner}(c).

%\textcolor{red}{The center localized mode in Fig.~\ref{fig:spring_mode}(a) as well as the mode in Fig.~\ref{fig:zero_main}(c) is similar to the corner mode shown in Fig.~\ref{fig:square_lat}(b) in the sense that all exhibit same wave decay pattern in the nontrivial segments. In the Fig.~\ref{fig:zero_main}(c) mode, The center nodes along diagonal $BB^\prime$ are connected only to the center nodes along the  $AA^\prime$ via springs of stiffness $k_i$. The equal and out-of-phase motion of the nodes along the diagonal $BB^\prime$ kepng the center nodes along the  $AA^\prime$ diagonal at rest. Now, if we consider a nonzero $k_1$ that is smaller than $k_2$ as in case of the center mode in Fig.~\ref{fig:spring_mode}(a), the center localized mode will decay along edges connected to oscillating center nodes. The decay pattern is same as shown in Fig.~\ref{fig:zero_corner}(c). Apart from the analogy, the frequency of the mode in Fig~\ref{fig:zero_main}(c) lies completely in the bandgap of the dispersion surface of the interface chain. The location of the frequency is marked by magenta dashed line in the dispersion diagram shown in Fig.~\ref{fig:zero_main}(e). Thus, the mode in Fig~\ref{fig:zero_main}(c) as well as the mode in Fig.~\ref{fig:spring_mode}(a) shows rapid wave decay along the interface chain.} 

\begin{figure}[!t]
    \centering
    \includegraphics[height = 10cm]{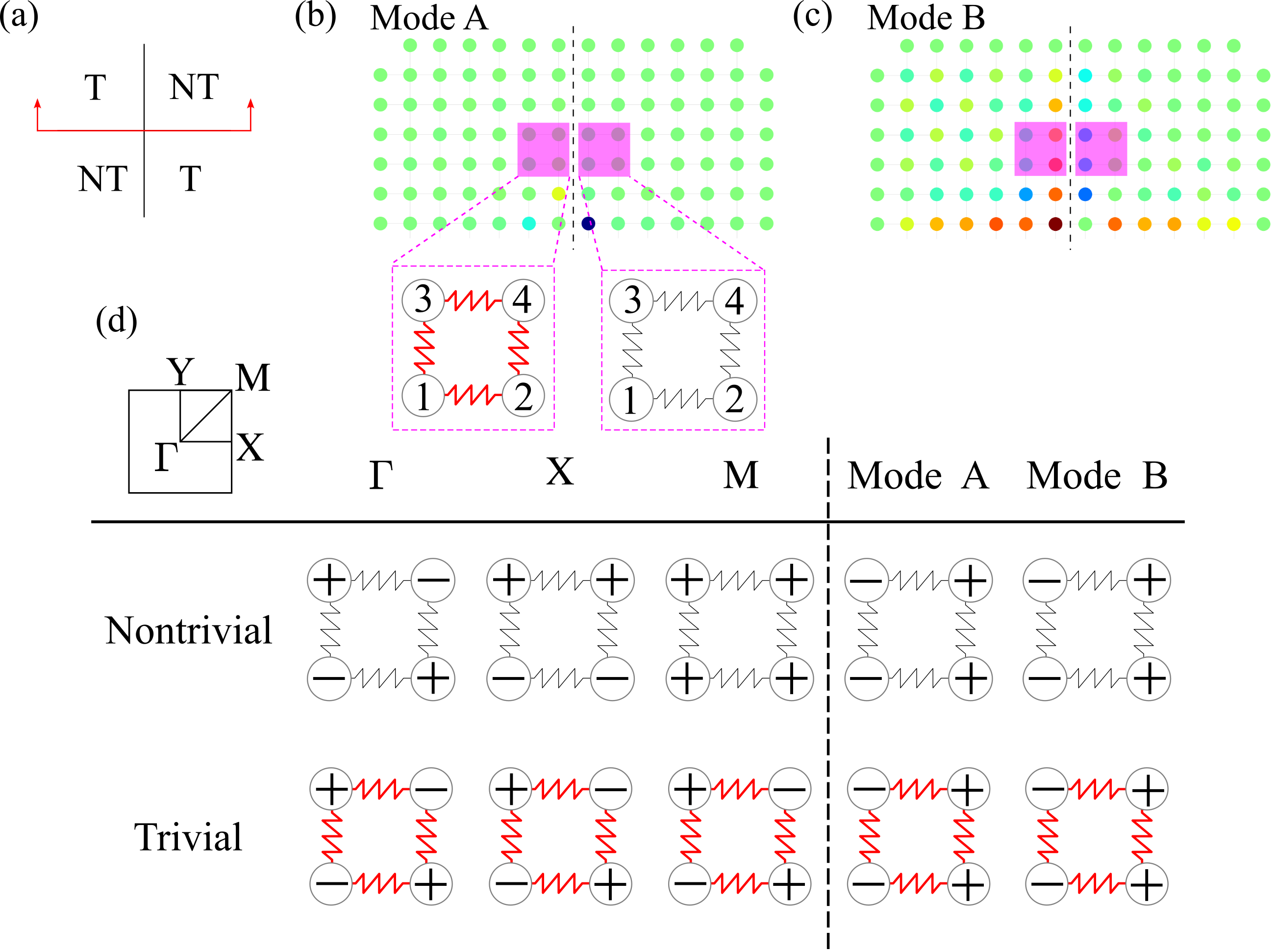}
    \caption{(a-c) Top half of the two center localized modes. The displacement in shaded unit cells are compared with Bloch mode shapes. (d) Displacement signs of the Bloch mode shapes at high symmetry points of the IBZ, along with the localized mode shape in the shaded unit cells in (b). The mismatch in symmetry causes the modes to decay away from the center.}
    \label{fig:ucMode}
\end{figure}

%Next, let us discuss the origin of the localized mode shown in Fig.~\ref{fig:spring_mode}(d). This mode has peak amplitude at the center nodes along the $AA^\prime$ diagonal. The center localized mode decays along the interface chain. Its frequency is marked by a green dashed line in the dispersion diagram in Fig.~\ref{fig:zero_main}(e). It lies just outside the pass band and thus results in a slowly decaying mode. If we lower the stiffness of four springs in the $FLF'L'$\textcolor{red}{\textbf{not sure how to write FLF'L'}} by an equal amount, the localized mode frequency decreases and will lie around the center of the bandgap between 4th and 5th frequency band. In this case,  the corresponding center localized mode decays rapidly away from the defect region. Now, instead of zero $k_1$, the lattice with $k_1 = 0.3076$ supports a similar mode decaying along the interface shown in Fig.~\ref{fig:spring_mode}(d). 

\subsubsection{Symmetry mismatch between bulk and localized modes}\label{sec:dis_symmetry_mismatch} 

Apart from the interface, let us analyze why the wave is not propagating in trivial and nontrivial lattices in the localized mode shown in Fig.~\ref{fig:spring_mode}. Here, we show how a mismatch between the symmetry of the localized modes and that of the traveling Bloch modes in the interior causes the modes to decay in both the trivial and nontrivial lattices.

The frequency of both of the localized modes in Fig.~\ref{fig:spring_mode} lies on the fourth dispersion branch, \textit{i.e.,} the red branch in Fig.~\ref{fig:square_lat}(e)). Figure~\ref{fig:ucMode}(b,c) display the top half of center localized modes shown in Fig.~\ref{fig:spring_mode}(a,b), respectively, along with two (shaded) unit cells each in the trivial and nontrivial segments. We will compare the displacement fields in these unit cells with the Bloch modes in the fourth dispersion branch. Figure~\ref{fig:ucMode}(d) displays the Bloch mode shapes in a trivial and a nontrivial unit cell at the high symmetry points of the irreducible Brillouin zone. Note that due to symmetry, the magnitude of displacement at each the four masses is equal to the same value or is zero. Thus the mode shape is indicated by either their relative signs $\{+,-\}$ or $0$ at each node. The last two columns in Fig.~\ref{fig:ucMode}(d) display signs of the localized mode shapes in the shaded unit cell in Mode A and Mode B. 

Although we present the mismatch arguments only by comparison with the high symmetry points, they extend to all the Bloch modes of this dispersion branch. We make the following observations regarding the mode shape's symmetries: 

\begin{itemize}

    \item Along $X$: For waves traveling along $\pm x$ directions, we examine the displacement of masses along that boundary of a unit cell which has outward normal toward $\pm x$-direction. The two vertical boundaries have normals in the $\pm x$ direction. Let us examine the displacement of the two masses on each vertical boundary, $(1,3)$ and $(2,4)$, in the Bloch mode shapes along the $\Gamma X$ line. Figure~\ref{fig:ucMode}(d) displays the signs of these mode shapes at the ends $\Gamma$  and $X$. They are out-of-phase, \textit{i.e.,} $\{+,-\}$ or $\{-,+\}$ in both the trivial and nontrivial unit cells. The displacement field is thus anti-symmetric about the horizontal axis through the unit cell center. Thus for waves propagating along $\pm x$ direction in the frequency range of this dispersion branch, the vertical masses in each unit cell move out-of-phase. In contrast, for both the localized modes, the vertical masses move in-phase, \textit{i.e.,} $\{-,-\}$ or $\{+,+\}$ in both the trivial and nontrivial unit cells. 

    \item Along diagonals: The Bloch mode shapes along the $\Gamma M$ line on this dispersion branch are symmetric about both diagonals. The masses along both diagonals are in-phase, \textit{i.e.,} $\{-,-\}$ or $\{+,+\}$ in both the trivial and nontrivial unit cells. This symmetric displacement implies that for wave propagating in the diagonal direction through the lattice, the displacement of diagonal masses in each unit cell are  in-phase. In contrast, both the localized modes have a different symmetry. The displacement of masses along both the diagonals is out-of-phase, \textit{i.e.,} $\{+,-\}$ or $\{-,+\}$ in both the trivial and  nontrivial unit cells. 

    \item Along $Y$: Since the trivial and nontrivial unit cells are each four fold rotation symmetric ($C_4$), if the Bloch modes and localized mode shapes have symmetry mismatch along $x$, they will also have the same mismatch along $y$. The mode shapes at the high symmetry point $Y$ of the Brillouin zone are not presented explicitly in Fig.~\ref{fig:ucMode}(d). They are $90^{\circ}$ rotated copies of the mode shapes at $X$ due to $C_4$ symmetry of the unit cell. In particular, for both the localized modes, the horizontal masses (1 and 2 as well as 3 and 4) move in-phase, \textit{i.e.,} $\{-,-\}$ or $\{+,+\}$ in both the trivial and nontrivial unit cells. In contrast, these masses move out-of-phase for each mode along the $\Gamma Y$ line in this dispersion branch. 
\end{itemize}

Let us now express a mode shape arising due to a localized defect in the basis of mode shapes of a periodic lattice. The governing equation that the defect mode shape at frequency $\omega$ satisfies may be written as $(-\omega^2 \bm{M} + \bm{K})\bm{u}_d = \bm{0}$, with $\bm{M}$ and $\bm{K}$ being the mass and stiffness matrix of a large finite structure under periodic boundary conditions. The stiffness matrix can be decomposed into contributions from the periodic and defect parts as $\bm{K} = \bm{K}_{per} + \bm{K}_d$, which then leads to 
\begin{equation}\label{eqn:defect}
 (-\omega^2 \bm{M} + \bm{K}_{per})\bm{u}_d = -\bm{K}_{d}\bm{u}_d   . 
\end{equation}
Let us express the defect mode in the basis of mode shapes of the periodic lattice as $\bm{u}_d = \sum_{j} \alpha_j \bm{u}_j$. These mode shapes satisfy $\bm{K}_{per} \bm{u}_j= \omega_j^2 \bm{M} \bm{u}_j$, with $\omega_j$ being the natural frequency. Using orthogonality, Eqn.~\eqref{eqn:defect} leads to $\alpha_j = (\bm{u}_j, \bm{K}_{d}\bm{u}_d)/ (\omega^2 - \omega_j^2)$. As we have periodic boundary conditions, the mode shape $\bm{u}_j$ at frequency $\omega_j$ is a linear superposition of Bloch modes at that frequency. Due to the above symmetry mismatch $(\bm{u}_j, \bm{K}_{d}\bm{u}_d)=0$ and thus $\alpha_j = 0$ for all those modes lying on the same dispersion branch as the localized mode frequency $\omega$. 

Finally, we discuss why this $\alpha_j=0$ leads to the decay of a mode shape away from the defect. At a localized mode frequency $\omega$, only the Bloch modes at that frequency support traveling waves. Now, if to the contrary, the mode does not decay in the far field, it will imply the presence of a Bloch mode at frequency $\omega$ that is not orthogonal to the defect mode shape in the defect core. Since all the Bloch modes are orthogonal to $\bm{K}_d \bm{u}_d$, we infer the decay away from the defect. In summary, due to symmetry mismatch between the localized modes and the Bloch mode shapes in both the trivial and nontrivial lattices, the modes stay localized at the center as shown in Fig.~\ref{fig:spring_mode}. 

\section{Numerical simulations and experiments on architected plates}\label{sec:Numerical_and_exp}

In this section, we present the design of an architected plate based on the spring mass model discussed in Sec.~\ref{sec:lat_int}. Using $3D$ finite element simulations, we predict the existence of center quasi-BICs. These predictions of quasi-BICs are followed by their experimental observation using laser Doppler vibrometry on a fabricated sample. 

\subsection{Proposed design of architected plate}\label{sec:3d_design} 

\begin{figure}[!t]
    \centering
    \includegraphics[width=16cm]{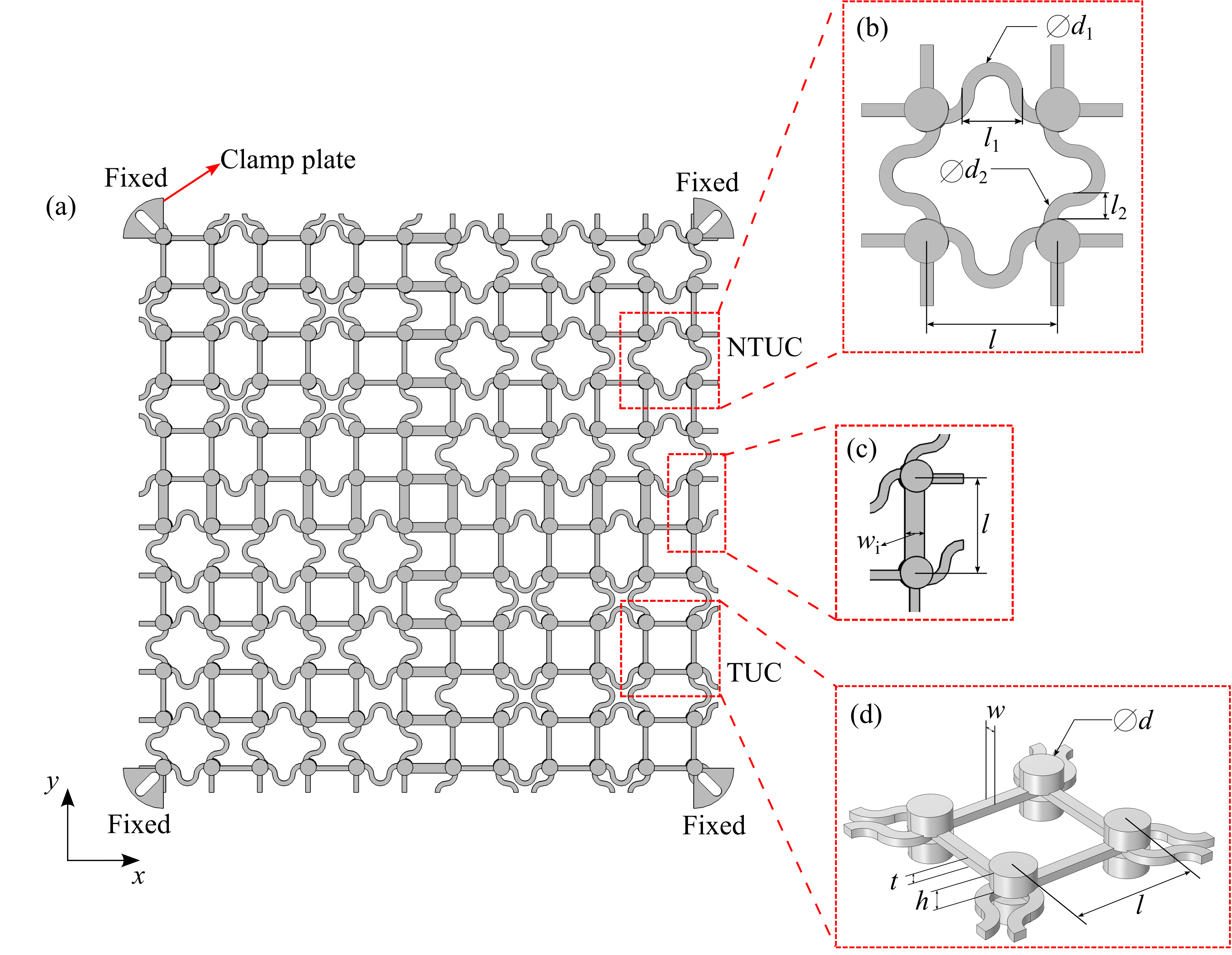}
    \caption{(a) Architected plate that is a continuous analogue of the discrete lattice in Fig~\ref{fig:spring_lat}(b). Enlarged views of (b) a nontrivial (c) an interface beam and (d) a trivial unit cell. Curved beams have lower stiffness than straight ones. Feature dimensions in the curved beam in (b) are $(d_a,d_b, l_a, l_b) = (14.08,8.08,14.08,6.46)$ mm.
    } 
    \label{fig:3d_lat}
\end{figure}

Figure~\ref{fig:3d_lat}(a) displays the three-dimensional ($3D$) model of an architected plate that is a continuous elastic analogue of the discrete system  in Fig.~\ref{fig:spring_lat}(b). It comprises of trivial and nontrivial parts with interfaces between them. Slender beams in the plate mimic springs in the discrete system. Similarly, cylindrical magnets placed at the lattice nodes are analogous to the point masses. Here, nodes refer to junctions where multiple slender beams meet. The magnets are placed on both sides of the plate so that their mutual attraction kepng them in place, as illustrated  for a unit cell in Fig.~\ref{fig:3d_lat}(d). The magnets at all the nodes are identical similar to identical masses in the discrete system. The distance between adjacent nodes is kept fixed as indicated by $l$ in Fig.~\ref{fig:3d_lat}, while the geometry of beam segments between the nodes is varied to achieve the stiffness distribution of the discrete system in Fig.~\ref{fig:spring_lat}(b).

\begin{figure}[!t]
    \centering
    \includegraphics[width = 15cm]{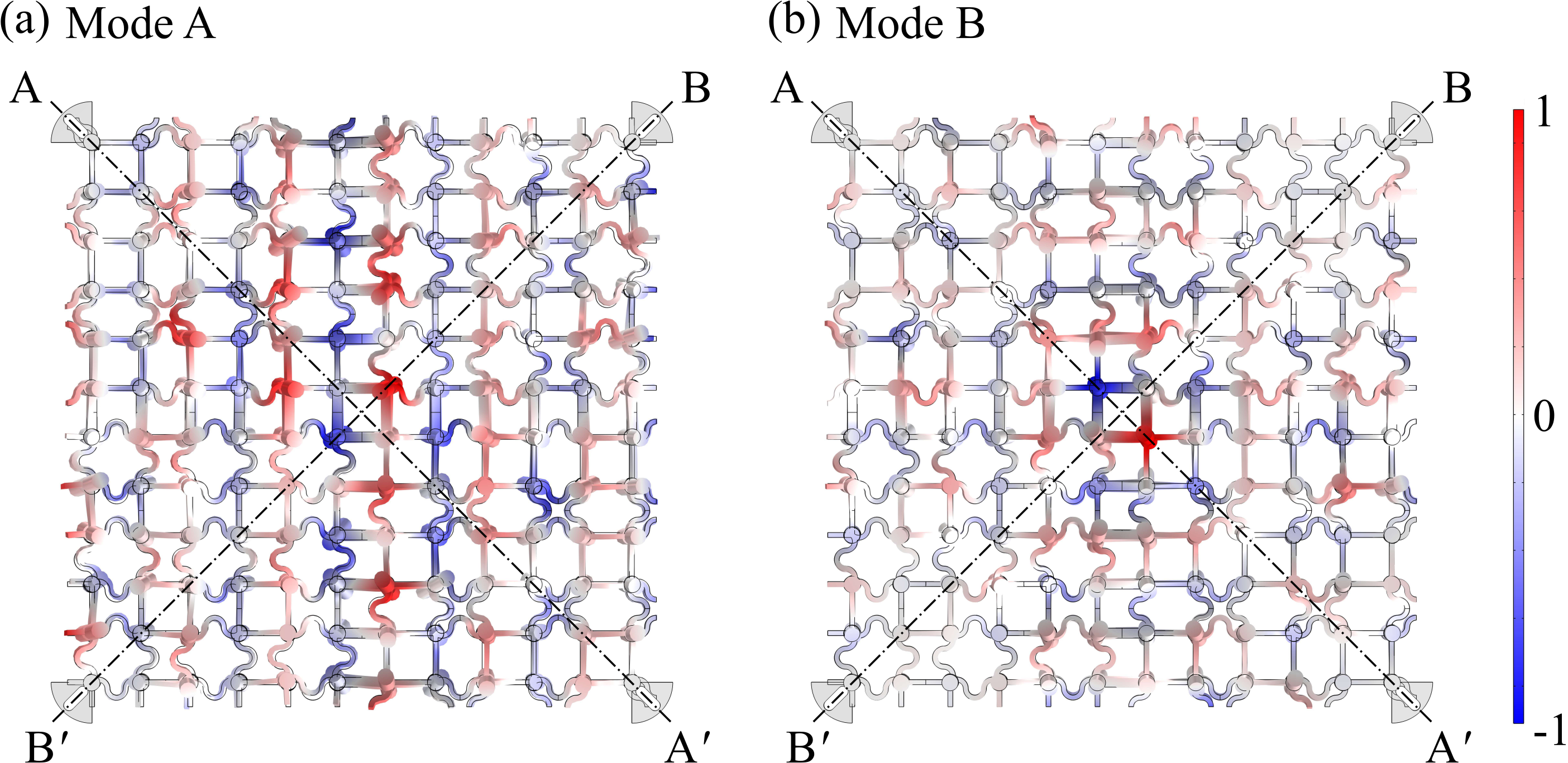}
    \caption{Center quasi-BICs in the plate determined by FEA. The modes in (a) and (b) are analogous to the mode of discrete lattice Fig.~\ref{fig:spring_mode}(a) and~\ref{fig:spring_mode}(b). Colorbar quantifies the out of plane displacement.}
    \label{fig:3d_mode}
\end{figure}

Let us discuss the beam design that results in an equivalent bending stiffness distribution as in the discrete system. Note that the bending stiffness of a beam with Young's modulus $E$, bending moment of inertia $I$ and length $L$ scales as $EI/L^3$~\cite{den2012strength}. For a rectangular cross-section with width $w$ and height $h$, $I = w h^3/12$. The height and Young's modulus are fixed, while the length and width of beams are varied across the plate. The stiffness scales in terms of these variables as $w/L^3$. To reduce the stiffness of a beam, it is made curved thereby increasing its length, while to increase stiffness, its width is increased. 

Figures~\ref{fig:3d_lat}(b),\ref{fig:3d_lat}(c) and \ref{fig:3d_lat}(d) display enlarged views of a nontrivial unit cell, an interface beam between trivial and nontrivial segments and a trivial unit cell, respectively. The curved and straight  beam segments are analogous to springs $k_1$ and $k_2$, respectively, with the latter having a higher stiffness due to its shorter length. The straight beams each have a length $l_2 = 27$ mm. Each curved beam has a total length $l_1 = 40$ mm, with its relevant dimensions presented in the caption in  Fig.~\ref{fig:3d_lat}(c). Both beams have the same width. The bending  stiffness ratio of a curved beam to a straight beam is thus $(EI/l_1^3)/(EI/l_2^3) = (l_2/l_1)^3 = 0.3076$, which is the same as that in the discrete system in Sec.~\ref{sec:theory}. 
Next, let us discuss the beams at the interface between trivial and nontrivial segments in the plate. The length of every interface beam is 27 mm, which is same as of the straight beam segment. To maintain same stiffness ratio $k_i/k_2 = 2$ as in the discrete model, the width of these interface beams are double that of the straight beams. The beam width $w$ is 3 mm except at the interfaces between the four lattices, where it is $w_i = 2w = 6$ mm.   

The plate is fixed at its four corners by a suitably designed small clamping segment. This clamp plate is a part of the structure as shown in Fig.~\ref{fig:3d_lat}(a) and makes it easy to fix the structure in the experimental setup. We will only focus on the center localized modes, which should not get perturbed based on conditions imposed at the far-field boundaries. 

\begin{figure}[!b]
     \centering    
     \includegraphics[width=14cm]{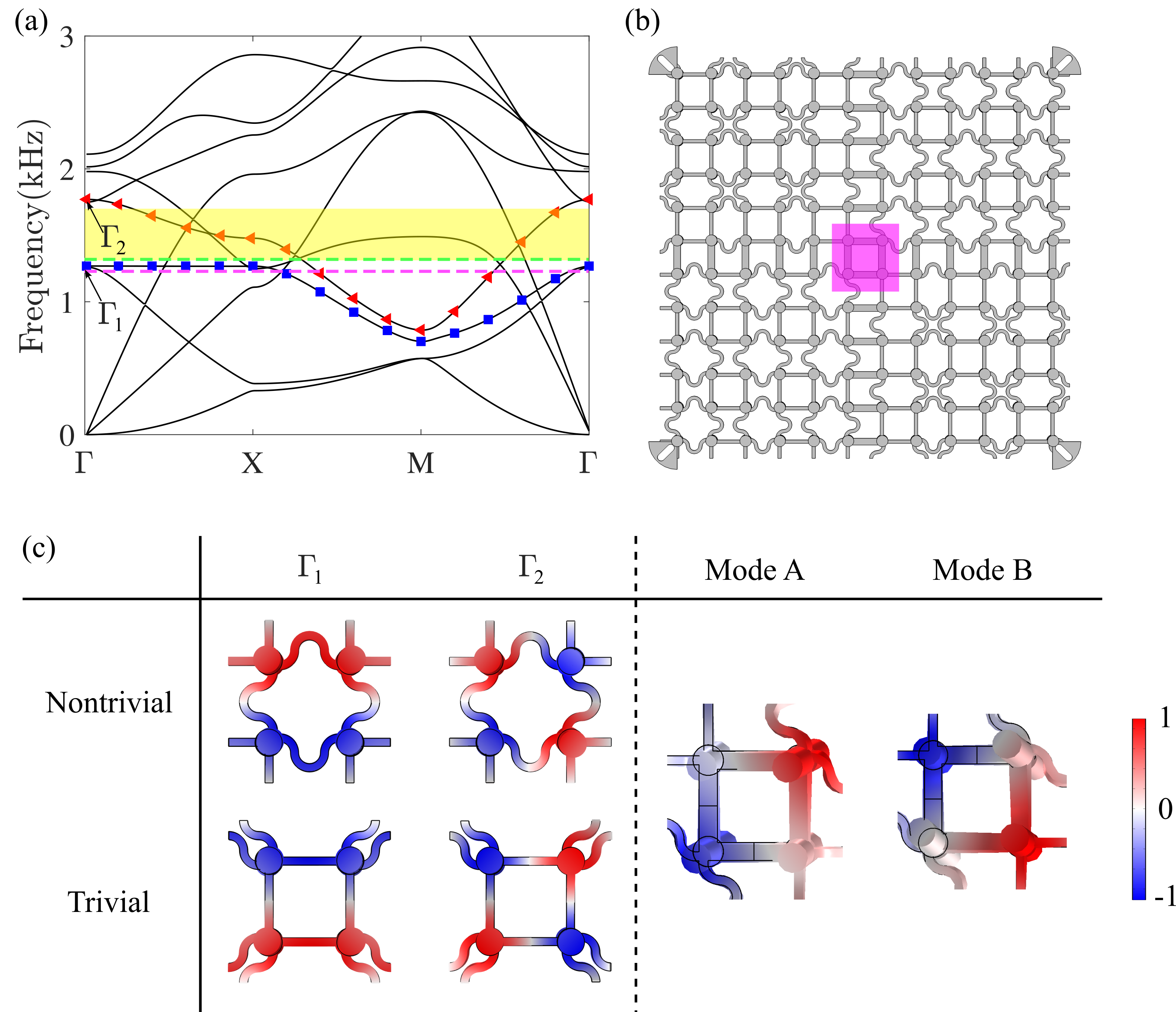}
     \caption{(a) Dispersion curves of plate lattice along the IBZ boundary. Center localized frequencies marked by dashed lines lie on the bending branches indicated by markers. (b) Bloch mode shapes of the trivial and nontrivial unit cells in the bending branch, along with  corresponding displacement fields of defect center in the two localized modes.  } 
     \label{fig:3d_disp}
\end{figure}

\subsection{Numerical simulations to identify center quasi-BICs}\label{sec:3d_sim}

To determine the mode shapes of the architected plate, we use three-dimensional ($3D$) finite element analysis (FEA). The plate is assumed to be an isotropic linear elastic solid with governing equation $\rho\ddot{\bm{u}} - [(\lambda + \mu)\nabla(\nabla\cdot\bm{u})+\mu\nabla^2\bm{u}] = 0$. Here $\bm{u} = (u_x,u_y,u_z)$ is the vector of displacement components, $\lambda$ and $\mu$ are the Lam\'e constants, and $\rho$ is the density of the solid~\cite{achenbach2012wave}. The plate is discretized using 1362159 quadratic tetrahedral elements and COMSOL MULTIPHYSICS is used to perform the FEA. The material properties for the plate are taken to be aluminum 6061 (Young's modulus $E = 68.9$ GPa, Poisson's ratio $\nu = 0.3$, density $\rho = 2700$ kg/m$^3$). For the nodes, neodymium magnet N35 (density $\rho_m =  7537.6$ kg/m$^3$) properties are used. The aluminum plate has constant thickness $t = 2.032$ mm (0.08 inch). The height ($h$) and diameter ($d$) of the magnets are 4.6 mm and 5 mm, respectively. The entire domain of the clamp plates are kept fixed, while the rest of the boundary is traction free in simulations.

Figure~\ref{fig:3d_mode} displays the center-localized modes in the plate. The diagonals $AA^\prime$ and $BB^\prime$ help us compare the mode shapes of these localized modes with those from the discrete model. The plate mode shape in Fig.~\ref{fig:3d_mode}(a) is analogous to the discrete lattice mode shape in Fig.~\ref{fig:spring_mode}(a). In both modes, the center masses along the $BB^\prime$ diagonal are moving out-of-phase and  those along $AA^\prime$ diagonal have zero displacement. Likewise, the mode shown in Fig.~\ref{fig:3d_mode}(b) is analogous to the discrete system's mode shape shown in Fig.~\ref{fig:spring_mode}(b). The frequencies of the mode in Fig.~\ref{fig:3d_mode}(a) and~\ref{fig:3d_mode}(b) are 1229 Hz and 1319 Hz, respectively. 

To determine if these localized modes are indeed quasi-BICs, we computed the dispersion curves of the nontrivial lattice using its unit cell. Figure~\ref{fig:3d_disp}(a) displays the dispersion curves along the path $\Gamma X M\Gamma$ connecting the high symmetry points in the irreducible Brillouin zone of this lattice. The magenta and green dashed lines indicate the frequencies of the two localized modes shown in Fig.~\ref{fig:3d_mode}(a) and~\ref{fig:3d_mode}(b), respectively. Examining the mode shapes at various frequencies along the  branch having red triangle markers reveals that these modes are flexural with predominantly out-of-plane displacement. As both the localized mode frequencies intersect this flexural mode branch, the localized modes are indeed quasi-BICs. 

Let us compare the symmetries of the localized modes with the corresponding Bloch modes as done earlier in  section~\ref{sec:dis_symmetry_mismatch} to understand their behavior away from the center. First, we analyze the localized mode labeled as ``Mode B" at $1319$ Hz. Its frequency is indicated by the green dashed line in Fig.~\ref{fig:3d_disp}(a) and this frequency intersects with multiple dispersion branches. An examination of the mode shapes in these dispersion branches reveals that only the red branch (with triangle markers) supports out-of-plane bending modes. The other dispersion branches that intersect with this localized mode frequency support only the in-plane modes. Figure~\ref{fig:3d_disp}(c) displays the Bloch mode shapes of a trivial and a nontrivial unit cell at the $\Gamma$ point, with the color code indicating the value of out-of-plane displacement $u_z$. In both these unit cells, $u_z$ is symmetric about each diagonal. The Bloch modes at the other high symmetry points $M,X$ of the IBZ also have the same symmetry, \textit{i.e.,} $u_z$ is symmetric about the diagonals. Hence we compare these Bloch mode shapes to the displacement field of the localized mode in the defect center, indicated by the shaded region in Fig.~\ref{fig:3d_disp}(b). Mode B is antisymmetric about both diagonals, showing a mismatch in symmetry with the Bloch modes. This mismatch results in its decay away from the defect center. 

Next, let us analyze the other localized mode labeled as ``Mode A" with frequency at $1229$ Hz. This frequency, marked by magenta dashed line, lies on two dispersion branches with flexural modes. The first is the same as discussed above, with red triangle markers. The second branch is marked by blue square markers in Fig.~\ref{fig:3d_disp}(a). The $u_z$ field in the Bloch modes at $\Gamma$ point of this dispersion branch are anti-symmetric about their diagonals in both the trivial and nontrivial unit cells. The localized mode A has the same symmetry as this branch at the defect center, \textit{i.e.}, $u_z$ is anti-symmetric about diagonals. Thus localized mode A hybridizes with Bloch modes with frequencies on blue square marked branch. As a result, mode A does not decay but oscillates away from the defect center, \textit{i.e.,} it is more leaky than mode B.

Based on the above analysis of Bloch mode shapes, we identify the yellow shaded region in Fig.~\ref{fig:3d_disp}(a) where only the red dotted branch supports out-of-plane modes. If a center localized mode frequency lies in this region, it will not hybridize with the bulk modes due to symmetry mismatch, thus resulting in its strong localization. 
If the frequency of the mode in Fig.~\ref{fig:3d_mode}(a) can be moved into the yellow shaded region in Fig.~\ref{fig:3d_disp}(a), the mode be strongly confined. One possible approach is to change the design of interface beams in the architected plate. Note that the interface beam design dictates the frequency of the localized mode. Making it stiffer (softer), this frequency will increase (decrease). 
Our work provides a systematic approach to realize localized modes in pass band frequencies. It involves comparing the mode shapes of the resulting modes at the defect core with the Bloch modes of dispersion branches on which the localized mode lies.

\subsection{Experimental observation of a quasi-BIC}\label{sec:exp}

\begin{figure}[!t]
    \centering
    \includegraphics[width=10cm]{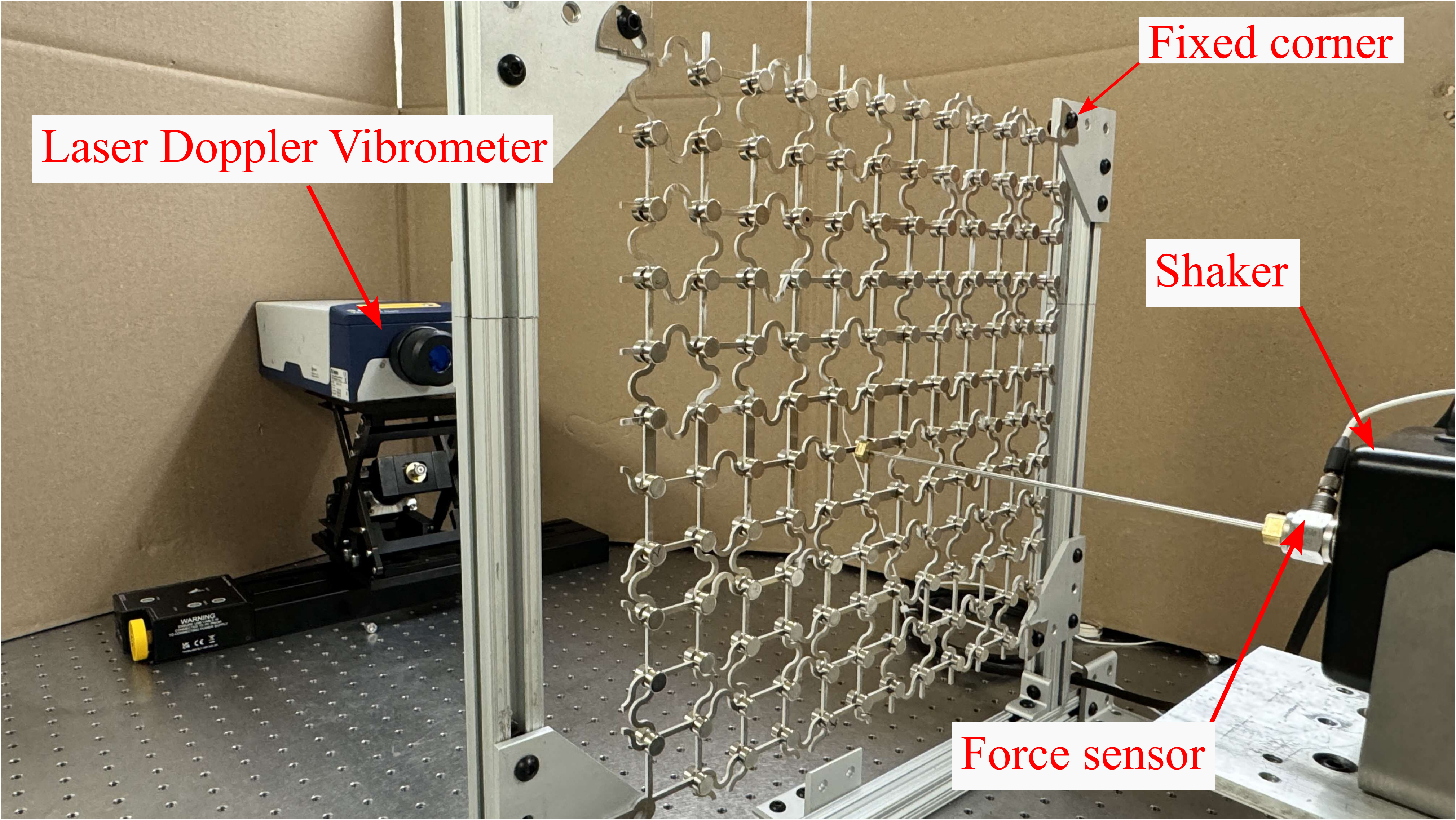}   
    \caption{Experimental setup to observe a center quasi-BIC in the plate. The plate is clamped at its corners, excited by a shaker and its velocity at nodes is measured with a laser vibrometer.}
    \label{fig:exp_set}
\end{figure}

\begin{figure}[!t]
    \centering
    \includegraphics[width=10cm]{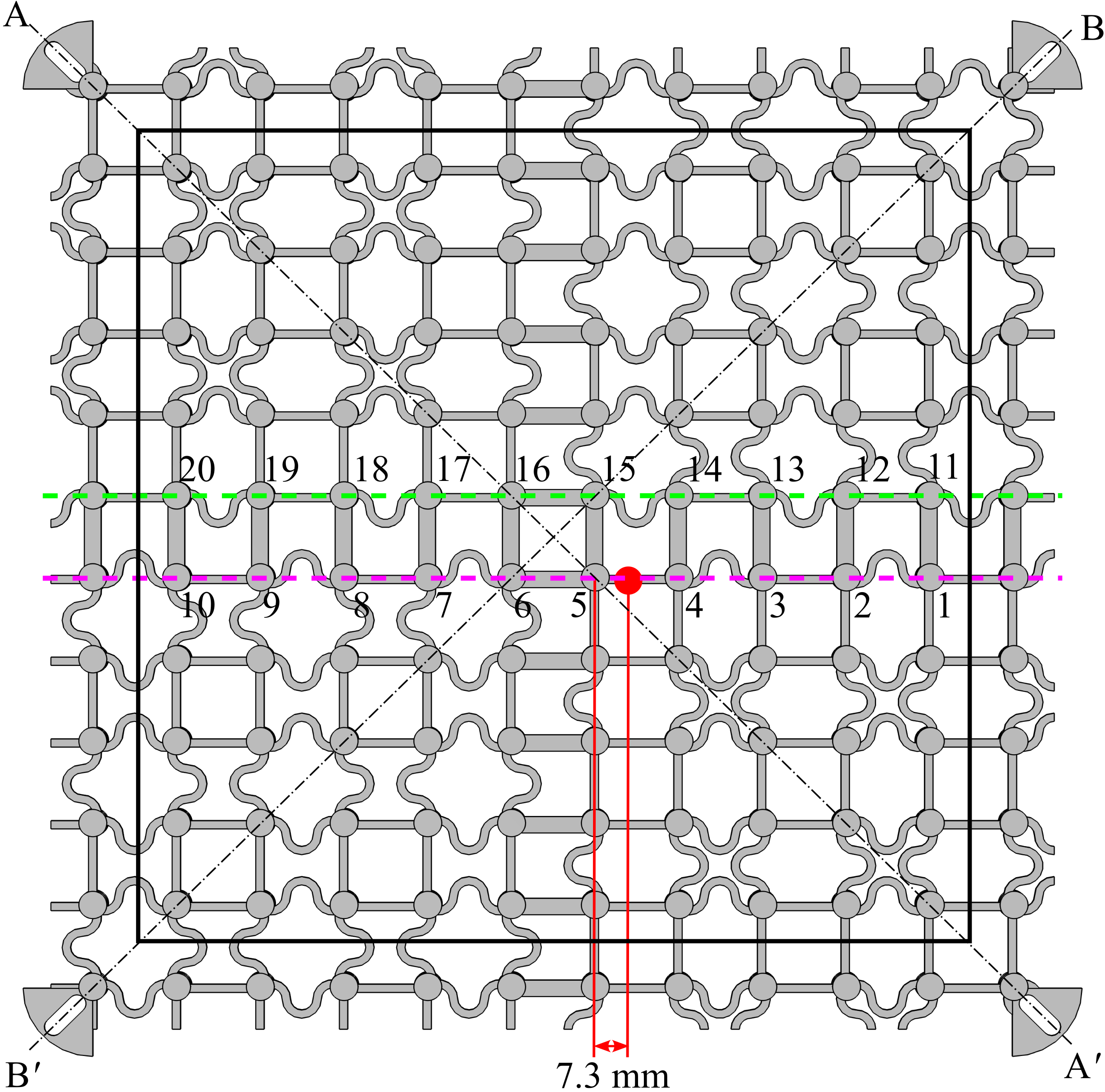}   
    \caption{Excitation and measurement locations in the plate. It is excited at the red dot location. The frequency response is measured at the labeled nodes lying on the two dashed lines. The mode shapes are extracted at the resonant frequencies by measuring the velocity at nodes enclosed by the black square. }
    \label{fig:exp_plate}
\end{figure}

Finally, we conducted experiments to observe the quasi-BIC presented in Fig.~\ref{fig:3d_mode}(b). The setup is shown in Fig.~\ref{fig:exp_set}(a). It is important to mention that, we also attempted to observe the localized mode at 1229 Hz shown in Fig.~\ref{fig:3d_mode}(a), but it was hard to observe that mode experimentally, as there is another bulk flexural mode at a close by frequency, 1239 Hz. We fabricated the architected structure from a 0.08 inch (2.032 mm) thick aluminum 6061 plate by water jet cutting. Commercially available cylindrical neodymium N35 magnets of 5 mm diameter and 4.6 mm height are placed at the top and bottom of each node. The plate is clamped at its four corners using screws through the slot in clamp plate. A permanent magnet shaker (Labworks ET-132-203) is used to excite the plate. The excitation point, indicated by the red circle in Fig.~\ref{fig:exp_plate}, is at a beam segment located 7.3 mm away from the center of the node along diagonal $AA^\prime$. A force sensor (PCB Piezotronics 208C01) is attached to the shaker to determine the applied force and a laser Doppler vibrometer (Polytec VFX-I-110) is used to measure velocity at different point on the architected plate. The vibrometer is placed on horizontal and vertical translation stages, so that it can be moved simultaneously along both directions. 

The experimental procedure involves two stpng. First we determine the frequency response of the plate, and seek for the resonance peak which corresponds to the center localized mode shown in Fig.~\ref{fig:3d_mode}(b), whose frequency is predicted 1319 Hz by $3D$ FEA. We excite the plate for different frequencies in the range from 1300 to 1350 Hz, and determine the frequency response for 20 magnets along the magenta and green dashed line shown in Fig.~\ref{fig:exp_plate}. The measured magnets are labeled by corresponding numbers in Fig.~\ref{fig:exp_plate}. For every frequency in the test frequency range, the plate is excited for 100 ms and the velocity of a particular magnet and force applied by shaker are recorded for last 5 ms.  This procedure ensures that the transients die out and we obtain the steady state response at the excitation frequency $f$. Now, by Fourier transformation of the measured velocity and force, the maximum velocity $v$ and  maximum applied force $F$ amplitudes at  this frequency  are determined. Then the normalized kinetic energy of a magnet at frequency $f$ is determined as $|v/F|^2$. 

\begin{figure}[!t]
    \centering
\includegraphics[width=16cm]{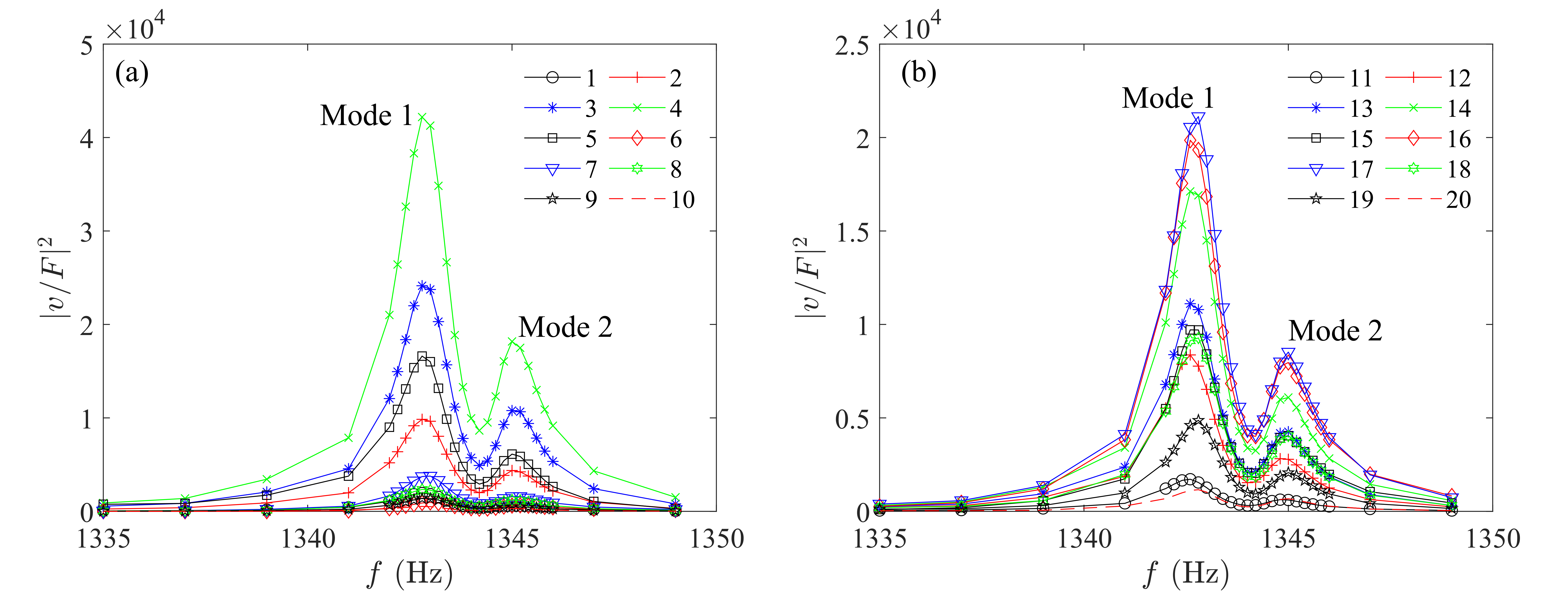}
    \caption{(a) Measured frequency response of the labeled nodes along the (a) magenta and (b) green dashed lines in Fig.~\ref{fig:exp_plate}. There are two resonant peaks: `Mode 1' and `Mode 2', in this frequency range. The velocity decreases away from the center in both these modes. }
    \label{fig:exp_freq}
\end{figure}

After exciting the plate at a frequency for 100 ms, the plate is kept at rest for one second to damp out any residual vibrations. Then the plate is excited again for 100 ms at the next frequency in test frequency range. This procedure is repeated for all the labeled nodes over the considered frequency range. Figures~\ref{fig:exp_freq}(a,b)  display the frequency response for these nodes. Results are presented for frequencies in the range 1335 Hz to 1350 Hz, with markers indicating the measured frequencies. We observe the presence of two resonance peaks at 1342 and 1345 Hz, and label them as Mode 1 and Mode 2. We also did measurements for frequencies between 1300 to 1335 Hz, but did not find any resonant peaks in this range.  

\begin{figure}[!b]
    \centering
\includegraphics[width=18cm]{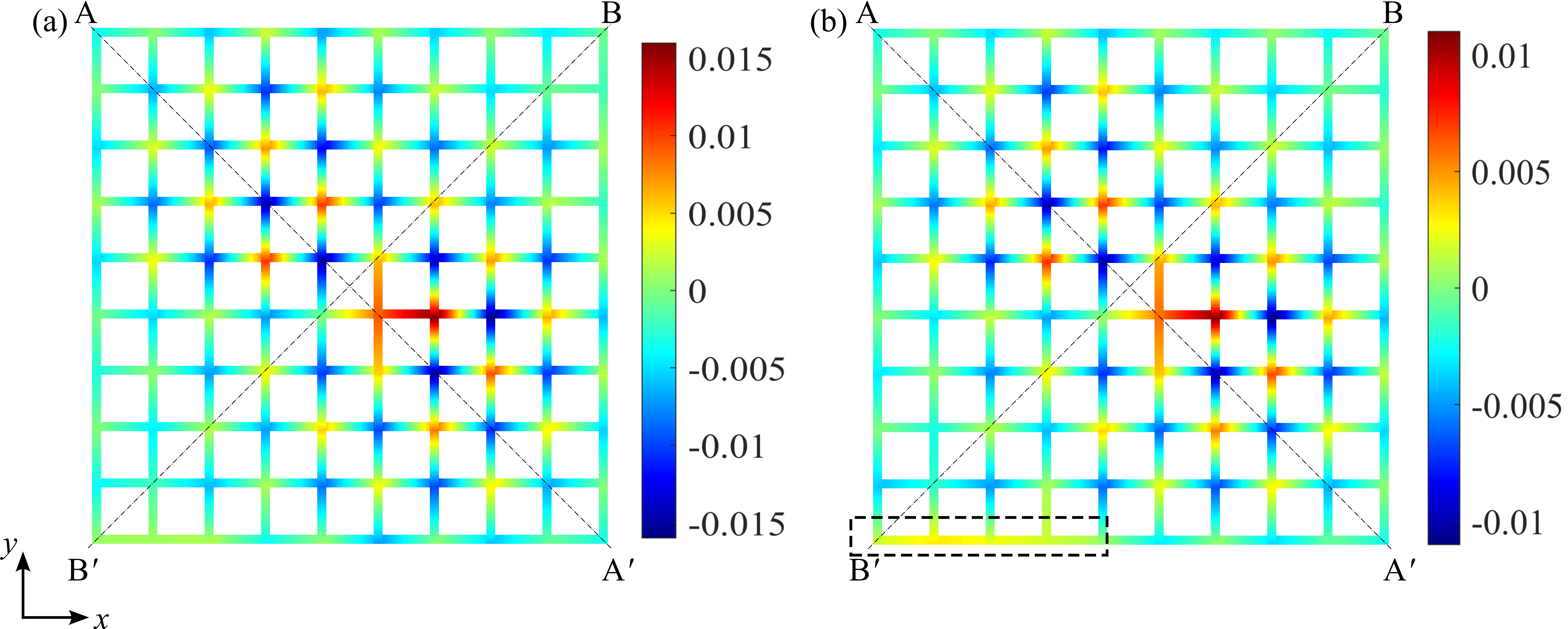}
    \caption{Measured velocity field at the two peak frequencies shown in Fig.~\ref{fig:exp_freq}: (a) `Mode 1' and (b) `Mode 2'. Color bar indicates out-of-plane velocity.}
    \label{fig:exp_mode}
\end{figure}

Our next step involves extracting the mode shapes experimentally for the two resonant peaks in Fig.~\ref{fig:exp_freq}(a,b). To this end, we measure the velocity vs time $v_p(t)$ for each node $p$ lying within the black box in Fig.~\ref{fig:exp_plate}. 
For a mode with frequency $\omega_0$, we determine the Fourier transform $\tilde{v}_p(\omega)$ of the measured velocity at each node. The absolute value of $\tilde{v}_p(\omega_0)$ gives the magnitude of velocity at node $p$ in the mode shape, while the phase is obtained by comparing the phase of $\tilde{v}_p(\omega_0)$
with the corresponding value at a reference node, node 5 here. 

The velocity field in the beam segments is rendered using a linear interpolation between the adjoining nodes. Figures~\ref{fig:exp_mode}(a,b) display the velocity fields in the enclosed domain at frequencies 1342 and 1345 Hz. At both frequencies, there is a  significant localization at the center of the plate. At the second resonant peak frequency, the lower left edge of the plate also has a high velocity. This additional peak at $1345$ Hz is due to a bulk mode, whose frequency from FEA simulations is 1335 Hz. Its mode shape is shown in Fig~\ref{fig:bulk}, see Appendix~\ref{sec:bulk} for details. Recall that the center localized mode frequency was $1319$ Hz from FEA. In the experiment, the second resonant peak corresponds to a mode that is a result of the corner localized mode hybridizing with this bulk mode. As a result, we observe a higher velocity amplitude just outside the center defect and the mode shown in Fig.~\ref{fig:exp_mode}(b) has a high velocity field in the region marked by black dashed line. Our experiments thus illustrate the existence of center localized modes in the pass band frequencies in the architected plate.

\section{Conclusion} \label{sec:con}
We investigated the dynamics of architected plates and reported the existence of center localized quasi-BICs in it. Our design is inspired by a topologically nontrivial square lattice that supports electronic BICs at its corners. Its discrete mechanical analogue, a finite square lattice has corner localized modes when it is comprised of nontrivial unit cells, \textit{i.e.,} when the boundary is appropriately terminated. However the presence of multiple bulk modes at the same frequency makes it hard to observe the corner localized modes. This degeneracy is broken by introducing a lattice that comprises of two copies each of trivial and nontrivial square lattices. The resulting lattice supports two center localized modes in the passband frequencies. We show that these localized modes decay away from the center due to their symmetry mismatch with propagating Bloch modes at that frequency. 

Building on the above insights, an architected plate is designed with a bending stiffness variation similar to that in the discrete lattice. The stiffness variation in the plate is achieved by using distinct types of beams: curved ones for lower stiffness and wider ones for higher stiffness. The architected plate is fabricated by removing materials from a single monolithic plate, showing our approach to be versatile to realize various wave phenomena. 

Using finite element analysis, we find two center localized modes in this plate with similar mode shapes as in the discrete lattice. Compared to the discrete lattice, the plate supports additional deformation modes and thus has a higher number of dispersion branches. One of the localized modes rapidly decays away from the center since its frequency lies on a dispersion branch whose Bloch modes have a symmetry mismatch with the localized mode shape. In contrast, the other localized mode lies on two dispersion branches that support flexural/bending modes. The mode shape has the same symmetry as the Bloch modes of one of these branches. It does not decay, but it instead oscillates with a small amplitude away from the center. However, moving the natural frequency of this mode out of the frequency range of this other dispersion branch by modifying the design of beams at the interface leads to its stronger confinement. The theoretical and numerical predictions are validated by experimental measurements of the velocity field in a fabricated structure. Our work thus shows a systematic way to confine energy by having a symmetry mismatch with Bloch modes of dispersion branches having the localized mode frequency. 

\section*{Acknowledgements}
This work was supported by the U.S. National Science
Foundation under Award No. 2503614. 

\appendix

\section{Natural frequencies and mode shapes of a finite discrete lattice}\label{sec:nat_freq}
Here we outline the procedure to determine the natural frequencies and corresponding mode shapes of a discrete spring mass lattice consisting of $N$ masses. Let us denote displacement of a node with index $p$ by $u_p$. The governing equation of motion of the mass at node $p$ is 
\begin{equation*}
    m\ddot{u}_p + \sum_{q = 1}^N k_{pq}\left(u_p - u_q\right) = F_p, 
\end{equation*}
where $k_{pq}$ is the stiffness of spring connecting nodes $p$ and $q$, and it is zero if $p$ and $q$ are not connected. $F_p$ is the external force applied to node $p$. To determine natural frequencies and mode shapes, we set $F_p = 0$ for all nodes and set the displacement to zero for nodes that are fixed.  The governing equations for the remaining free nodes in lattice may be written in the matrix form as $\bm{M}\ddot{\bm{u}} + \bm{Ku} = \bm{0}$. We seek solutions in the form $\bm{u}(t) =  \bm{\tilde{u}}e^{i\omega t}$, which yields an eigenvalue problem $\omega^2 \bm{M\tilde{u}} = \bm{K\tilde{u}}$. Its solution yields the set of natural frequencies $\omega$ and corresponding mode shapes $\bm{\tilde{u}}$. The modes are classified as localized or bulk by plotting and visually examining the displacement field. 

\section{Dispersion analysis of unit cell}\label{sec:disp_analysis}
To determine the dispersion surface of a nontrivial unit cell, let us consider a unit cell with indices $(p,q)$ in the $(x,y)$ directions in an infinite-square lattice. The governing equations for the masses in this unit cell are
\begin{align*}
m \ddot{u}_{p,q,1} + k_1 (u_{p,q,1} - u_{p,q,2})   + k_1(u_{p,q,1} - u_{p,q,3}) 
+ k_2 (u_{p,q,1} - u_{p-1,q,2})   + k_2 (u_{p,q-1,1} - u_{p,q,3}) &= 0 ,
\\ 
m \ddot{u}_{p,q,2} + k_1 (u_{p,q,2} - u_{p,q,1})   + k_1 (u_{p,q,2} - u_{p,q,4}) 
+ k_2 (u_{p,q,2} - u_{p+1,q,1})   + k_2 (u_{p,q,2} - u_{p,q-1,4}) &= 0 ,
\\ 
m \ddot{u}_{p,q,3} + k_1 (u_{p,q,3} - u_{p,q,4})   + k_1 (u_{p,q,3} - u_{p,q,1}) 
+ k_2 (u_{p,q,3} - u_{p-1,q,4})   + k_2 (u_{p,q,3} - u_{p+1,q,1}) &= 0 ,
\\ 
m \ddot{u}_{p,q,4} + k_1 (u_{p,q,4} - u_{p,q,3})   + k_1 (u_{p,q,4} - u_{p,q,2}) 
+ k_2 (u_{p,q,4} - u_{p+1,q,3})   + k_2 (u_{p,,4} - u_{p,q+1,2}) &= 0 .
\end{align*} 
Let $\bm{u}_{p,q}$ denote the vector containing displacements of the unit cell's nodes. 
Using Bloch-Floquet periodicity conditions, the displacement of any unit cell is expressed in terms of the reference unit cell as 
$\bm{u}_{p,q}(t) =  \bm{\hat{u}}e^{i(\omega t -  p\mu_x - q\mu_y)}$. The governing equation is turned into the form of an eigenvalue problem as $-\omega^2\bm{M} \bm{\hat{u}} + \bm{K} \bm{\hat{u}} = 0$. Here, $\mu_x$ and $\mu_y$ are non-dimensional wavenumbers in $x$ and $y$ direction, respectively. As we consider the nodal masses of unit mass, the mass matrix $\bm{M}_{p,q}$ is a $4\times4$ identity matrix. The stiffness matrix $\hat{\bm{K}}_{p,q}$ for a nontrivial unit cell is 
\begin{equation*}
    \bm{\hat{K}} = \begin{bmatrix}
        2(k_1+k_2) &-k_1-k_2e^{-i\mu_x} &-k_1-k_2e^{-i\mu_y} &0\\
        -k_1-k_2e^{i\mu_x} &2(k_1+k_2) &0 &-k_1-k_2e^{-i\mu_y}\\
        -k_1-k_2e^{i\mu_y} &0 &2(k_1+k_2) &-k_1-k_2e^{-i\mu_x}\\
        0 &-k_1-k_2e^{i\mu_y} &-k_1-k_2e^{i\mu_x} &2(k_1+k_2)\\
    \end{bmatrix}. 
\end{equation*}
Note that the dispersion surfaces of lattices with trivial and nontrivial unit cells are identical since they are the same (infinite) lattice. We solve the eigenvalue problem for each value of $\mu_x$ and $\mu_y$ in the first or irreducible Brillouin zone, as desired.
%\begin{comment}

\section{Effect of interface stiffness $k_i$ on center mode frequencies}\label{sec:centerModeVary}
\begin{figure}[!hbtp]
    \centering
    \includegraphics[height = 6cm]{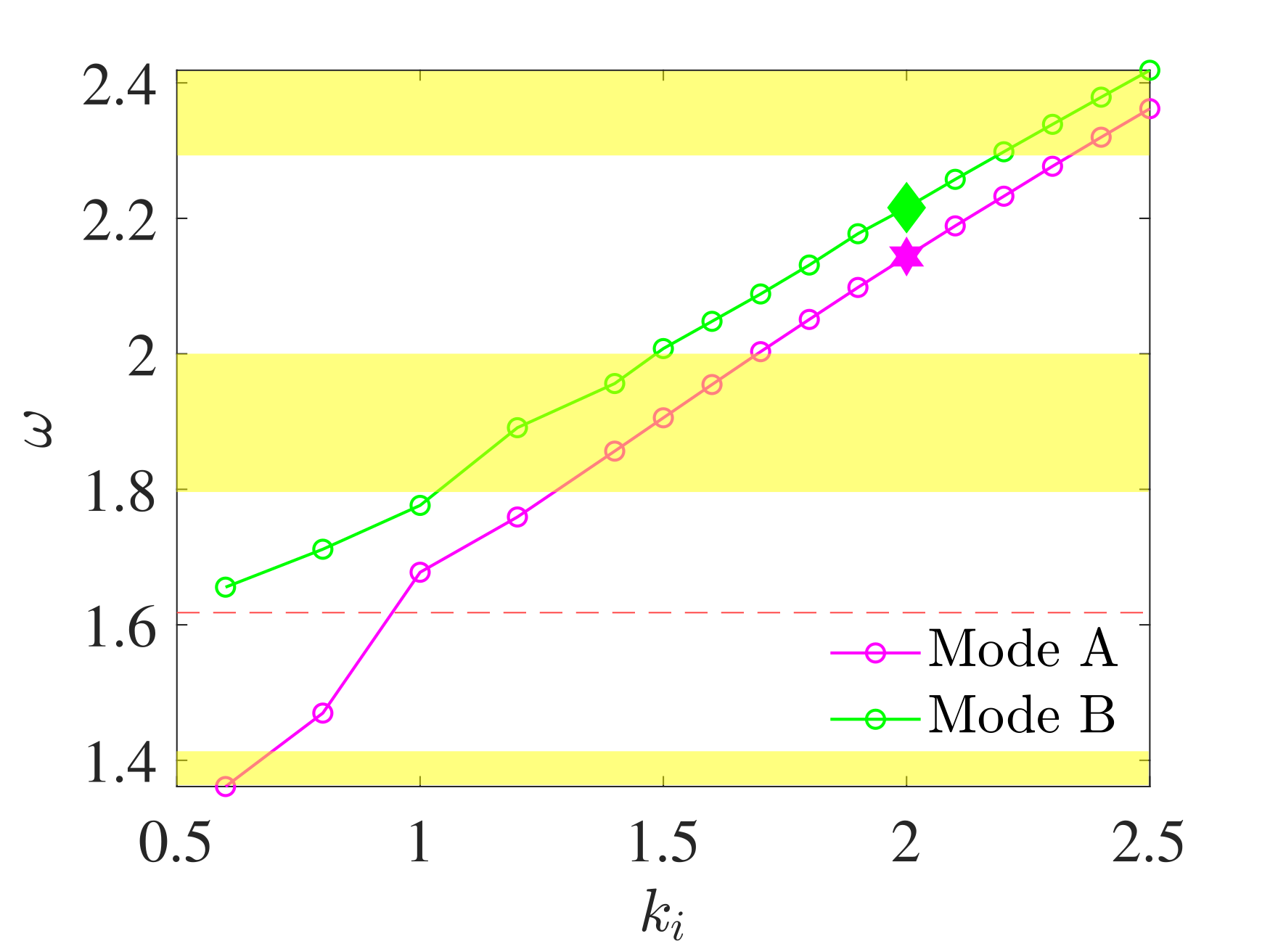}
    \caption{Variation of center localized mode frequencies as the interface spring stiffness $k_i$ is varied in the discrete lattice in Fig.~\ref{fig:spring_lat}(b). }
    \label{fig:app}
\end{figure}

Figure~\ref{fig:app} displays how the frequencies of center localized modes in Fig.~\ref{fig:spring_mode}(a) (Mode A) and~\ref{fig:spring_mode}(b) (Mode B) vary as a function of the interface spring stiffness, $k_i$ in the discrete system of Fig.~\ref{fig:spring_lat}(b). The yellow shaded regions indicate the bandgap regions in the dispersion surface (see Fig.~\ref{fig:square_lat}(e)) of the square lattice. The red dashed line marks the corner localized mode frequency. The magenta star and green diamond mark the center localized mode frequencies for $k_1 = 2$ which is considered in our proposed design. The figure shows that by varying stiffness at interface, the center localized modes' frequency can be significantly shifted across the dispersion surface. 

%\end{comment}
\section{Bulk mode shape close to center localized mode frequency}\label{sec:bulk}
\begin{figure}[!hbtp]
    \centering
    \includegraphics[width=0.5\linewidth]{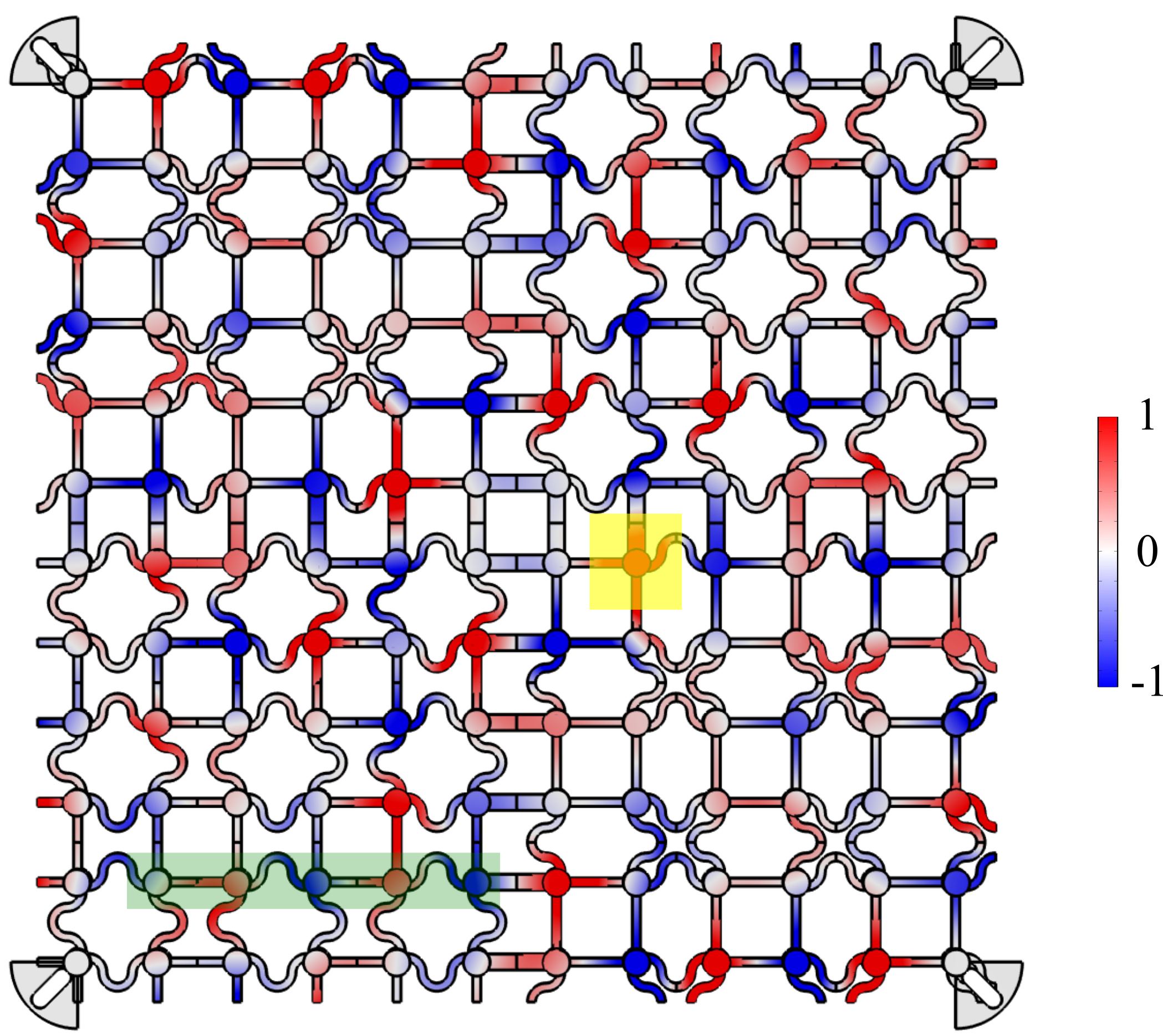}
    \caption{Bulk mode shape determined using FEA corresponding to the resonance peak frequency denoted as `Mode 2' in Fig.~\ref{fig:exp_freq}}
    \label{fig:bulk}
\end{figure}

We discuss the reason for the presence of two resonance peaks in the experiment. Figure~\ref{fig:bulk} displays a bulk mode whose frequency (1335 Hz) lies close to the center localized mode frequency (1319 Hz) shown in Fig.~\ref{fig:3d_mode}b. Both this localized mode and the bulk mode have maximum amplitude close to the excitation point. In particular, the node enclosed by the yellow shaded box in Fig.~\ref{fig:bulk} has the maximum amplitude. 
Thus, both center and bulk mode are excited in the experiment. We infer that the second resonance peak arises due to the bulk mode by observing (a) a high velocity field in the region marked by black dashed line in Fig.~\ref{fig:exp_mode}(b) in the experimental result and (b) displacement field in the corresponding region, green shaded box in Fig:~\ref{fig:bulk},  from FEA.

\bibliography{main}

\begin{thebibliography}{10}

\bibitem{hussein2014dynamics}
Mahmoud~I Hussein, Michael~J Leamy, and Massimo Ruzzene.
\newblock Dynamics of phononic materials and structures: Historical origins, recent progress, and future outlook.
\newblock {\em Applied Mechanics Reviews}, 66(4):040802, 2014.

\bibitem{hernandez2008localized}
Hugo~E Hern{\'a}ndez-Figueroa, Michel Zamboni-Rached, and Erasmo Recami.
\newblock {\em Localized waves}, volume 194.
\newblock John Wiley \& Sons, 2008.

\bibitem{zhu2014negative}
Rui Zhu, XN~Liu, GK~Hu, CT~Sun, and GL~Huang.
\newblock Negative refraction of elastic waves at the deep-subwavelength scale in a single-phase metamaterial.
\newblock {\em Nature communications}, 5(1):5510, 2014.

\bibitem{zanotto2022metamaterial}
Simone Zanotto, Giorgio Biasiol, Paulo~V Santos, and Alessandro Pitanti.
\newblock Metamaterial-enabled asymmetric negative refraction of ghz mechanical waves.
\newblock {\em Nature Communications}, 13(1):5939, 2022.

\bibitem{li2012elastic}
FengMing Li and YiZe Wang.
\newblock Elastic wave propagation and localization in band gap materials: a review.
\newblock {\em Science China Physics, Mechanics and Astronomy}, 55:1734--1746, 2012.

\bibitem{jui2024topological}
Tamanna~Akter Jui and Raj~Kumar Pal.
\newblock Topological localized modes in moir{\'e} lattices of bilayer elastic plates with resonators.
\newblock {\em Journal of Sound and Vibration}, 576:118268, 2024.

\bibitem{hsu2016bound}
Chia~Wei Hsu, Bo~Zhen, A~Douglas Stone, John~D Joannopoulos, and Marin Solja{\v{c}}i{\'c}.
\newblock Bound states in the continuum.
\newblock {\em Nature Reviews Materials}, 1(9):1--13, 2016.

\bibitem{xu2023recent}
Guizhen Xu, Hongyang Xing, Zhanqiang Xue, Dan Lu, Jinying Fan, Junxing Fan, Perry~Ping Shum, and Longqing Cong.
\newblock Recent advances and perspective of photonic bound states in the continuum.
\newblock {\em Ultrafast Science}, 3:0033, 2023.

\bibitem{huang2022general}
Lujun Huang, Bin Jia, Artem~S Pilipchuk, Yankei Chiang, Sibo Huang, Junfei Li, Chen Shen, Evgeny~N Bulgakov, Fu~Deng, David~A Powell, et~al.
\newblock General framework of bound states in the continuum in an open acoustic resonator.
\newblock {\em Physical Review Applied}, 18(5):054021, 2022.

\bibitem{imada1999coherent}
Masahiro Imada, Susumu Noda, Alongkarn Chutinan, Takashi Tokuda, Michio Murata, and Goro Sasaki.
\newblock Coherent two-dimensional lasing action in surface-emitting laser with triangular-lattice photonic crystal structure.
\newblock {\em Applied physics letters}, 75(3):316--318, 1999.

\bibitem{hirose2014watt}
Kazuyoshi Hirose, Yong Liang, Yoshitaka Kurosaka, Akiyoshi Watanabe, Takahiro Sugiyama, and Susumu Noda.
\newblock Watt-class high-power, high-beam-quality photonic-crystal lasers.
\newblock {\em Nature photonics}, 8(5):406--411, 2014.

\bibitem{kodigala2017lasing}
Ashok Kodigala, Thomas Lepetit, Qing Gu, Babak Bahari, Yeshaiahu Fainman, and Boubacar Kant{\'e}.
\newblock Lasing action from photonic bound states in continuum.
\newblock {\em Nature}, 541(7636):196--199, 2017.

\bibitem{Romano:19}
Silvia Romano, Gianluigi Zito, Sof\'{i}a N.~Lara Y\'{e}pez, Stefano Cabrini, Erika Penzo, Giuseppe Coppola, Ivo Rendina, and Vito Mocellaark.
\newblock Tuning the exponential sensitivity of a bound-state-in-continuum optical sensor.
\newblock {\em Opt. Express}, 27(13):18776--18786, Jun 2019.

\bibitem{Doskolovich:19}
Leonid~L. Doskolovich, Evgeni~A. Bezus, and Dmitry~A. Bykov.
\newblock Integrated flat-top reflection filters operating near bound states in the continuum.
\newblock {\em Photon. Res.}, 7(11):1314--1322, Nov 2019.

\bibitem{foley2014symmetry}
Justin~M Foley, Steven~M Young, and Jamie~D Phillips.
\newblock Symmetry-protected mode coupling near normal incidence for narrow-band transmission filtering in a dielectric grating.
\newblock {\em Physical Review B}, 89(16):165111, 2014.

\bibitem{kawachi2001optimal}
Osamu Kawachi, Seiji Mineyoshi, Gou Endoh, Masanori Ueda, Osamu Ikata, E~Hashimoto, and Masatsune Yamaguchi.
\newblock Optimal cut for leaky saw on litao/sub 3/for high performance resonators and filters.
\newblock {\em IEEE transactions on ultrasonics, ferroelectrics, and frequency control}, 48(5):1442--1448, 2001.

\bibitem{naumenko2003surface}
Natalya~F Naumenko and Benjamin~P Abbott.
\newblock Surface acoustic wave devices using optimized cuts of a piezoelectric substrate, April~29 2003.
\newblock US Patent 6,556,104.

\bibitem{cao2021perfect}
Liyun Cao, Yifan Zhu, Sheng Wan, Yi~Zeng, Yong Li, and Badreddine Assouar.
\newblock Perfect absorption of flexural waves induced by bound state in the continuum.
\newblock {\em Extreme Mechanics Letters}, 47:101364, 2021.

\bibitem{benabid2002stimulated}
Fetah Benabid, Jonathan~C Knight, G~Antonopoulos, and P~St~J Russell.
\newblock Stimulated raman scattering in hydrogen-filled hollow-core photonic crystal fiber.
\newblock {\em Science}, 298(5592):399--402, 2002.

\bibitem{couny2007generation}
F~Couny, Fetah Benabid, PJ~Roberts, PS~Light, and MG~Raymer.
\newblock Generation and photonic guidance of multi-octave optical-frequency combs.
\newblock {\em Science}, 318(5853):1118--1121, 2007.

\bibitem{rahman2022bound}
Adib Rahman and Raj~Kumar Pal.
\newblock Bound modes in the continuum based phononic waveguides.
\newblock {\em Journal of Applied Physics}, 132(11):115109, 2022.

\bibitem{PhysRevApplied.19.054001}
Bin Jia, Lujun Huang, Artem~S. Pilipchuk, Sibo Huang, Chen Shen, Almas~F. Sadreev, Yong Li, and Andrey~E. Miroshnichenko.
\newblock Bound states in the continuum protected by reduced symmetry of three-dimensional open acoustic resonators.
\newblock {\em Phys. Rev. Appl.}, 19:054001, May 2023.

\bibitem{cong2019symmetry}
Longqing Cong and Ranjan Singh.
\newblock Symmetry-protected dual bound states in the continuum in metamaterials.
\newblock {\em Advanced Optical Materials}, 7(13):1900383, 2019.

\bibitem{PhysRevA.100.063803}
Shiyu Li, Chaobiao Zhou, Tingting Liu, and Shuyuan Xiao.
\newblock Symmetry-protected bound states in the continuum supported by all-dielectric metasurfaces.
\newblock {\em Phys. Rev. A}, 100:063803, Dec 2019.

\bibitem{sadrieva2019experimental}
ZF~Sadrieva, MA~Belyakov, MA~Balezin, PV~Kapitanova, EA~Nenasheva, AF~Sadreev, and AA~Bogdanov.
\newblock Experimental observation of a symmetry-protected bound state in the continuum in a chain of dielectric disks.
\newblock {\em Physical Review A}, 99(5):053804, 2019.

\bibitem{plotnik2011experimental}
Yonatan Plotnik, Or~Peleg, Felix Dreisow, Matthias Heinrich, Stefan Nolte, Alexander Szameit, and Mordechai Segev.
\newblock Experimental observation of optical bound states in the continuum.
\newblock {\em Physical review letters}, 107(18):183901, 2011.

\bibitem{PhysRevLett.100.183902}
D.~C. Marinica, A.~G. Borisov, and S.~V. Shabanov.
\newblock Bound states in the continuum in photonics.
\newblock {\em Phys. Rev. Lett.}, 100:183902, May 2008.

\bibitem{sadreev2021interference}
Almas~F Sadreev.
\newblock Interference traps waves in an open system: bound states in the continuum.
\newblock {\em Reports on Progress in Physics}, 84(5):055901, 2021.

\bibitem{bulgakov2008bound}
Evgeny~N Bulgakov and Almas~F Sadreev.
\newblock Bound states in the continuum in photonic waveguides inspired by defects.
\newblock {\em Physical Review B}, 78(7):075105, 2008.

\bibitem{marinica2008bound}
DC~Marinica, AG~Borisov, and SV~Shabanov.
\newblock Bound states in the continuum in photonics.
\newblock {\em Physical review letters}, 100(18):183902, 2008.

\bibitem{vaidya2021point}
Sachin Vaidya, Wladimir~A Benalcazar, Alexander Cerjan, and Mikael~C Rechtsman.
\newblock Point-defect-localized bound states in the continuum in photonic crystals and structured fibers.
\newblock {\em Physical review letters}, 127(2):023605, 2021.

\bibitem{zhen2014topological}
Bo~Zhen, Chia~Wei Hsu, Ling Lu, A~Douglas Stone, and Marin Solja{\v{c}}i{\'c}.
\newblock Topological nature of optical bound states in the continuum.
\newblock {\em Physical review letters}, 113(25):257401, 2014.

\bibitem{benalcazar2020bound}
Wladimir~A Benalcazar and Alexander Cerjan.
\newblock Bound states in the continuum of higher-order topological insulators.
\newblock {\em Physical Review B}, 101(16):161116, 2020.

\bibitem{cerjan2020observation}
Alexander Cerjan, Marius J{\"u}rgensen, Wladimir~A Benalcazar, Sebabrata Mukherjee, and Mikael~C Rechtsman.
\newblock Observation of a higher-order topological bound state in the continuum.
\newblock {\em Physical review letters}, 125(21):213901, 2020.

\bibitem{wang2021quantum}
Yao Wang, Bi-Ye Xie, Yong-Heng Lu, Yi-Jun Chang, Hong-Fei Wang, Jun Gao, Zhi-Qiang Jiao, Zhen Feng, Xiao-Yun Xu, Feng Mei, et~al.
\newblock Quantum superposition demonstrated higher-order topological bound states in the continuum.
\newblock {\em Light: Science \& Applications}, 10(1):173, 2021.

\bibitem{zhang2023super}
Zhanyuan Zhang, Evgeny Bulgakov, Konstantin Pichugin, Almas Sadreev, Yi~Xu, and Yuwen Qin.
\newblock Super quasibound state in the continuum.
\newblock {\em Physical Review Applied}, 20(1):L011003, 2023.

\bibitem{farhat2024observation}
Mohamed Farhat, Younes Achaoui, Julio Andr{\'e}s~Iglesias Mart{\'\i}nez, Mahmoud Addouche, Ying Wu, and Abdelkrim Khelif.
\newblock Observation of ultra-high-q resonators in the ultrasound via bound states in the continuum.
\newblock {\em Advanced Science}, 11(33):2402917, 2024.

\bibitem{overvig2021chiral}
Adam Overvig, Nanfang Yu, and Andrea Al{\`u}.
\newblock Chiral quasi-bound states in the continuum.
\newblock {\em Physical Review Letters}, 126(7):073001, 2021.

\bibitem{huang2022moire}
Lei Huang, Weixuan Zhang, and Xiangdong Zhang.
\newblock Moir{\'e} quasibound states in the continuum.
\newblock {\em Physical Review Letters}, 128(25):253901, 2022.

\bibitem{vial2024platonic}
Benjamin Vial, Marc~Mart{\'\i} Sabat{\'e}, Richard Wiltshaw, S{\'e}bastien Guenneau, and Richard~V Craster.
\newblock Platonic quasi-normal modes expansion.
\newblock {\em arXiv preprint arXiv:2407.12042}, 2024.

\bibitem{rahman2024elastic}
Adib Rahman and Raj~Kumar Pal.
\newblock Elastic bound modes in the continuum in architected beams.
\newblock {\em Physical Review Applied}, 21(2):024002, 2024.

\bibitem{haq2021bound}
Omer Haq and Sergei Shabanov.
\newblock Bound states in the continuum in elasticity.
\newblock {\em Wave Motion}, 103:102718, 2021.

\bibitem{cao2021elastic}
Liyun Cao, Yifan Zhu, Yanlong Xu, Shi-Wang Fan, Zhichun Yang, and Badreddine Assouar.
\newblock Elastic bound state in the continuum with perfect mode conversion.
\newblock {\em Journal of the Mechanics and Physics of Solids}, 154:104502, 2021.

\bibitem{an2024multibranch}
Shuowei An, Tuo Liu, Liyun Cao, Zhongming Gu, Haiyan Fan, Yi~Zeng, Li~Cheng, Jie Zhu, and Badreddine Assouar.
\newblock Multibranch elastic bound states in the continuum.
\newblock {\em Physical Review Letters}, 132(18):187202, 2024.

\bibitem{fan2022observation}
Haiyan Fan, He~Gao, Shuowei An, Zhongming Gu, Yafeng Chen, Sibo Huang, Shanjun Liang, Jie Zhu, Tuo Liu, and Zhongqing Su.
\newblock Observation of non-hermiticity-induced topological edge states in the continuum in a trimerized elastic lattice.
\newblock {\em Physical Review B}, 106(18):L180302, 2022.

\bibitem{gao2024bound}
Nan Gao, Ricardo~Martin Abraham-Ekeroth, and Daniel Torrent.
\newblock Bound states in the continuum for antisymmetric lamb modes in composite plates made of isotropic materials.
\newblock {\em Wave Motion}, 129:103348, 2024.

\bibitem{fan2019elastic}
Haiyan Fan, Baizhan Xia, Liang Tong, Shengjie Zheng, and Dejie Yu.
\newblock Elastic higher-order topological insulator with topologically protected corner states.
\newblock {\em Physical review letters}, 122(20):204301, 2019.

\bibitem{zheng2022higher}
Zhoufu Zheng, Jianfei Yin, Jihong Wen, and Dianlong Yu.
\newblock Higher-order topological states in locally resonant elastic metamaterials.
\newblock {\em Applied Physics Letters}, 120(14), 2022.

\bibitem{chen2021corner}
Chun-Wei Chen, Rajesh Chaunsali, Johan Christensen, Georgios Theocharis, and Jinkyu Yang.
\newblock Corner states in a second-order mechanical topological insulator.
\newblock {\em Communications Materials}, 2(1):62, 2021.

\bibitem{PhysRevB.100.075120}
Ze-Guo Chen, Changqing Xu, Rasha Al~Jahdali, Jun Mei, and Ying Wu.
\newblock Corner states in a second-order acoustic topological insulator as bound states in the continuum.
\newblock {\em Phys. Rev. B}, 100:075120, Aug 2019.

\bibitem{wu2023square}
Shi-Qiao Wu, Zhi-Kang Lin, Zhan Xiong, Bin Jiang, and Jian-Hua Jiang.
\newblock Square-root higher-order topology in rectangular-lattice acoustic metamaterials.
\newblock {\em Physical Review Applied}, 19(2):024023, 2023.

\bibitem{padavic2018topological}
Karmela Padavi{\'c}, Suraj~S Hegde, Wade DeGottardi, and Smitha Vishveshwara.
\newblock Topological phases, edge modes, and the hofstadter butterfly in coupled su-schrieffer-heeger systems.
\newblock {\em Physical Review B}, 98(2):024205, 2018.

\bibitem{den2012strength}
Jacob~Pieter Den~Hartog.
\newblock {\em Strength of materials}.
\newblock Courier Corporation, 2012.

\bibitem{achenbach2012wave}
Jan Achenbach.
\newblock {\em Wave propagation in elastic solids}.
\newblock Elsevier, 2012.

\end{thebibliography}
\bibliographystyle{unsrt}

\newpage

\end{document}